\documentclass{amsart}
\usepackage{amscd}

\def\br#1{
 \ifx#1<\gdef\Br##1>{\left<##1\right>}\else
 \ifx#1(\gdef\Br##1){\left(##1\right)}\else
 \ifx#1[\gdef\Br##1]{\left[##1\right]}\else
 \ifx#1\{\gdef\Br##1\}{\left\{##1\right\}}\else
 \ifx#1|\gdef\Br##1|{\left|##1\right|}\else
 \ifx#1\|\gdef\Br##1\|{\left\|##1\right\|}\else
 \errmessage{MYMAC: Bad bracket!}\fi\fi\fi\fi\fi\fi
 \Br}
\def\seq{\subseteq}
\def\cj{\overline}
\def\ul{\underline}
\def\opn#1 {\operatorname{#1}}
\long\def\forgetit#1{\relax}
\newtheorem{thm}{Theorem}[subsection]
\newtheorem{cor}[thm]{Corollary}
\newtheorem{lem}[thm]{Lemma}
\newtheorem{prp}[thm]{Proposition}
\theoremstyle{remark}
\newtheorem*{rem}{Remark}
\newtheorem{rmn}[thm]{Remark}
\newtheorem*{rems}{Remarks}
\newtheorem{rmns}[thm]{Remarks}
\long\def\CAR#1#2\NIL{#1}
\long\def\Brm#1\Erm{
 \edef\nxt{\CAR#1\relax\NIL}
 \expandafter\ifx\nxt(
 \begin{rems} #1 \end{rems}\else
 \begin{rem} #1 \end{rem}\fi
}
\long\def\Brmn#1 #2\Erm{
 \edef\nxt{\CAR#2\relax\NIL}
 \expandafter\ifx\nxt(
 \begin{rmns}\mYlbl{#1} #2 \end{rmns}\else
 \begin{rmn}\mYlbl{#1} #2 \end{rmn}\fi
}
\def\mYlbl#1{\xdef\mYlastlabel{#1}\label{#1}}

\numberwithin{equation}{subsection}
\def\Beq#1\Eeq{\begin{equation*} #1 \end{equation*}}
\def\Beqn#1 #2\Eeq{\begin{equation}#2 \mYlbl{#1} \end{equation}}
\def\Bml#1\Eml{\begin{multline*} #1 \end{multline*}}
\def\Bmln#1 #2\Eml{\begin{multline}#2 \mYlbl{#1} \end{multline}}
\def\Bal#1\Eal{\begin{align*} #1 \end{align*}}
\def\Baln#1 #2\Eal{\begin{align}\mYlbl{#1} #2 \end{align}}
\def\Bcd#1\Ecd{\[\begin{CD} #1 \end{CD}\]}
\def\Bcdn#1 #2\Ecd{
 \begin{equation}\begin{CD}#2 \mYlbl{#1}\end{CD}\end{equation}}
\def\bysame{\leavevmode\hbox to3em{\hrulefill}\,}

\def\even{{\mathbf0}}
\def\odd{{\mathbf1}}
\def\seven{_\even}
\def\sodd{_\odd}
\def\sevR{_{\even,\mathbb R}}

\def\O{\mathop{\mathcal O}\nolimits}
\def\M{\mathop{\mathcal M}\nolimits}
\def\J{{\mathcal J}}
\def\Cau{^{\opn Cau }}
\def\CauV{^{\opn Cau,V }}
\def\V{^{\opn V }}
\def\D{^{\opn D }}
\def\sol{^{\opn sol }}
\def\rdmm{\mathbb R^{d+1}}
\def\rdm{\mathbb R^d}
\newdimen\mYd  \newbox\mYbox
\def\dcj#1{\def\mYarg{#1}
 \setbox\mYbox=\hbox{$\mYarg$}\mYd=\wd\mYbox
 \vbox{\offinterlineskip\hbox{\vrule width\mYd height1pt}\vskip1pt\box\mYbox}}
\def\causall{\prec}
\def\too{\xrightarrow}
\def\L{{\opn L }}
\def\Re{{\opn Re }}
\def\Im{{\opn Im }}
\def\supp{{\opn supp }}
\def\i{{\opn i }}
\def\Space{\opn space }
\def\free{^{\opn free }}
\def\kin{^{\opn kin }}
\def\bos{^{\opn bos }}
\def\trp{^{\opn T }}
\def\cX{{\mathcal X}}
\def\cD{{\mathcal D}}
\def\cE{{\mathcal E}}
\def\cEc{{\mathcal E}_{\opn c }}
\def\cH{{\mathcal H}}
\def\bV{{\mathbb V}}
\def\hA{\hat{\mathcal A}}
\def\hX{\hat{\mathcal X}}
\def\hP{\hat{\mathcal P}}
\def\cL{{\mathcal L}}
\def\LInt{{\mathcal V}}
\def\cA{{\mathcal A}}
\def\fo{^{\opn fo }}
\DeclareMathSymbol\square {\mathord}{AMSa}{"03} 
\def\KlGord{\square}
\def\Dirac{\not\!\! D}
\def\FUNref#1{\cite[#1]{[FUNCT]}}
\def\fg{{\mathfrak g}}
\def\exc{^{\opn exc }}
\def\param{^{\opn par }}
\def\extsol{^{\opn extsol }}
\def\cst{\opn cst }
\def\ext{^{\opn ext }}

\def\Erstr{E^0_1}
\def\scinf{_{C^\infty}}
\def\Bb#1#2{{}^{#1}_{#2}} 
\def\Ups{\Upsilon}

\begin{document}

\title[Supermanifolds of solutions]{
 Supermanifolds of classical solutions for Lagrangian \linebreak
 field models with ghost and fermion fields
}
\author{T. Schmitt}
\begin{abstract}
Using a supergeometric interpretation of field functionals, we show
that for a class of rather common classical field models used for realistic
quantum field theoretic models, an infinite-dimensional supermanifold
(smf) of classical solutions in Minkowski space can be constructed.
That is, we show that the smf of smooth Cauchy data with compact support
is isomorphic with an smf of corresponding classical solutions of the model.
\end{abstract}
\maketitle
\tableofcontents
\newpage

\section{Lagrangian field theories}

\subsection{Introduction}
This paper continues the investigation of the space of classical
solutions of geometric field models of quantum field theory started in
\cite{[FUNCT]}, \cite{[WHAT]}, \cite{[CAUCHY]}, \cite{[ABSEVEQ]}.
The interest in such spaces can be traced back to the sixties; cf.
\cite{[Segal60]}, \cite{[Segal74]}, \cite{[Sniatycki]}, \cite{[Sniatycki2]}.
The new feature in our investigations is the appropriate account of fermionic
degrees of freedom by using supergeometric methods, and the
actual mathematically rigorous construction of Poincar\'e invariant
solution spaces.

We start with fixing in \ref{ModClass} a class of classical
field models in Minkowski space $\rdmm$ which
contains a number of common models like e.~g. the $\Phi^4$ model,
the Thirring model, and the Yukawa model of meson-nucleon scattering.

Yang-Mills models have their peculiarities and will be treated separately,
in sections 3 and 4.

Also, $\sigma$ models as well as models with a non-linearized gravitational
field do not fit directly into our class; nevertheless,
our methods should be applicable also to them.

Using the framework of infinite-dimensional supergeometry constructed
in \cite{[FUNCT]}, \cite{[WHAT]}, and the
results on non-linear field equations with anticommuting fields
presented in \cite{[CAUCHY]}, \cite{[ABSEVEQ]}, we will solve
the classical field equations for compactly supported smooth Cauchy data.

While in purely bosonic models, like e.g. the $\Phi^4$ model (cf.
\cite{[CAUCHY]} for a discussion), one constructs a map
\Beqn NaiveXiSol
 \{\text{space of Cauchy data at $t=0$}\}\longrightarrow
 \{\text{space of configurations on space-time}\},
\Eeq
this is no longer appropriate in the presence of fermionic, anticommuting
fields, since both Cauchy data and
configurations form no longer sets (cf. \cite{[WHAT]} for a discussion of the
background).
Instead of this, they are described
by infinite-dimensional supermanifolds (smf's); thus, the correct
generalization
of \eqref{NaiveXiSol} is a morphism of supermanifolds
\Beqn RightXiSol
 \Xi\sol: M\Cau=\{\text{smf of Cauchy data at $t=0$}\}\longrightarrow
 \{\text{smf of configurations on space-time}\}=M.
\Eeq
A suitable calculus of infinite-dimensional smf's modelled on $\mathbb
Z_2$-graded locally convex vector spaces
has been constructed by the present author in \cite{[FUNCT]} and \cite{[WHAT]}
and will be used throughout in the rest of this paper.

Of course, both the map \eqref{NaiveXiSol} and the morphism
\eqref{RightXiSol} make sense only after suitable functional-analytic qualities
have been specified. A reasonable choice for the Cauchy data is the test
function space $\cEc\CauV:=\cD(\rdm)\otimes(V\oplus\dot V)$;
here $V$ is the target space for the fields, and
$\dot V$ is the target space for the velocities of those
fields the field equations of which are of order
two, i. e. the bosonic fields and the Faddeev-Popov ghosts, if these are
present.
For the configurations we take the space $\cEc\V$ of all those
$f\in C^\infty(\rdmm)\otimes V$ the support of which on every time slice is
compact and grows only with light velocity (cf. \ref{SpaceCic} for details).

Now we associate to a given model a {\em configuration supermanifold}, or
more precisely, the {\em supermanifold of smooth configurations with causally
growing spatially compact support}, which is the linear smf $\L(\cEc\V)$
(or "affine smf") modelled over the "naive configuration space" $\cEc\V$
\Beqn cEcSmf
 M = \L(\cEc\V)
\Eeq
with standard coordinate $\Xi\in\M^{\cEc\V}(M)$ (where $\M^{\cEc\V}(M)$
denotes the $\mathbb Z_2$-graded vector space of real, even $\cEc\V$-valued
superfunctions on $M$).
In particular, in the absence of fermionic fields, $M$ is simply the locally
convex space $\cEc\V$  viewed as infinite-dimensional manifold, and $\Xi$ is
the identity map $\cEc\V \to\cEc\V$.

Also, we need the  {\em supermanifold of compactly supported smooth Cauchy
data}
which is the linear smf
\Beqn cEcCauSmf
 M\Cau = \L(\cEc\CauV)
\Eeq
with standard coordinate $(\Xi\Cau,\dot\Phi\Cau)\in\M^{\cEc\CauV}(M\Cau)$,
where
$\dot\Phi\Cau$ are the initial velocities of the bosonic fields.
(Here and in the following, we suppose for notational simplicity that the
Pauli Theorem is valid, so that the second-order fields are exactly the
bosonic ones. The modifications in the presence of ghost fields are obvious;
cf.
section \ref{SecYMGBrGh}.)

For proceeding, we have to suppose that the underlying bosonic model
is {\em complete}, i. e. classically all-time solvable for smooth Cauchy data
with compact support. As to be expected, the existence of some a priori
estimate for the Sobolev norm of solutions of the bosonic field equations
is sufficient for completeness (cf. Thm. \ref{ComplThm}). We do not need
information on the continuous dependence of the solutions on the Cauchy
data since our approach automatically provides real-analytic dependence.

We will show that, as a consequence of our previous investigations
\cite{[CAUCHY]}, \cite{[ABSEVEQ]} of non-linear wave equations involving
anticommuting fields, there exists for a complete model a unique
$\cEc\V$-valued
superfunction on $M\Cau$,
\Beqn AnCauProbl
 \Xi\sol = \Xi\sol[\Xi\Cau,\dot\Phi\Cau]\in\M^{\cEc\V}(M\Cau),
\Eeq
which satisfies the field equations and has the correct Cauchy data:
\Beqn XiSolIsSol
 \frac\delta{\delta\ul\Xi_i}\cL[\Xi\sol]=0\quad
 \text{for $i=1,\dots,N$,}
\Eeq
\Beqn XiSolRightCD
 \Xi\sol[\Xi\Cau,\dot\Phi\Cau](0,\cdot) = \Xi\Cau(\cdot),\qquad
 \partial_t\Phi\sol[\Xi\Cau,\dot\Phi\Cau](0,\cdot) = \dot\Phi\Cau(\cdot).
\Eeq
The superfunction $\Xi\sol$ then determines the smf morphism \eqref{RightXiSol}
to be constructed. This morphism identifies $M\Cau$ with a sub-smf $M\sol$
of $M$ which we call the {\em supermanifold of classical solutions}. The
name is justified by the fact that given a morphism
$\phi:Z\to M$, i. e. a $Z$-family of
configurations, it factors through $M\sol$ iff we have
$\phi^*(\frac\delta{\delta\ul\Xi_i} \cL[\Xi]) =0$, i. e. $\phi$ solves the
field
equations.

In particular, the underlying manifold of $M\sol$ is just the set of all
those solutions (in the usual sense) of the underlying bosonic equations
which are smooth and compactly supported on every time slice.

If the Lagrangian is Poincar\'e invariant then
the natural action of the Poincar\'e group on $M$ restricts to
$M\sol$ and hence carries over onto $M\Cau$ (however, the formal proof
will be given only in the next part). In particular,
translation in time direction determines a complete flow (i. e.  a one
parameter automorphism group) on the smf $M\Cau$ which we call the
{\em time evolution flow}.

In \ref{CInfSolSmf}, we also consider smooth configurations and Cauchy data
without any support and growth condition, getting a morphism
$\Xi\sol: M\scinf\Cau\to M\scinf$ as a variant of \eqref{RightXiSol}.
(In fact, in the case of Yang-Mills theory in temporal gauge, this is the only
morphism we can construct since the arising constraint
\eqref{DaEaCst} is not sufficiently compatible with spatially compact support.)

Yet another variant arises by considering fluctuations around a fixed
bosonic "background" configuration which solves the bosonic field equations;
cf. \ref{LocExc}.

\subsection{A class of classical models}\label{ModClass}
Here we consider a class of Lagrangian field theories on Minkowski $\rdmm$
for which our approach immediately leads to uniqueness theorems, local
existence, and, if a non-blow-up of the local solutions can be guaranteed
from the outside, to an smf of solutions.

The reduction onto first-order equations which was assumed in
\cite{[CAUCHY]}, \cite{[ABSEVEQ]} is not natural in a Lorentz-invariant
context. Instead, we take the Lagrangians as they stand.

We will use $\mu,\nu,\dots{}=0,\dots,d$ as Lorentz indices which are raised
and lowered in the usual way: $T_\mu = g_{\mu\nu}T^\mu$,\ \
$(g_{\mu\nu}):=\opn diag (-1,1,\dots,1)$.
Also $a,b,\dots {} =1,\dots,d$ are spatial indices, and  $i,j$ will number
field components ranging from $1$ to $N^\Phi$ or  $N^\Psi$ or $N$ or \dots,
depending on the context. For all three types of indices, we will use the
Einstein sum convention.

We consider a field-theoretical model in $\rdmm=\mathbb R\times\rdm$ with
$N^\Phi$ real bosonic, commuting second-order fields
$\Phi_1,\dots,\Phi_{N^\Phi}$
(e. g. bosonic matter fields, or Yang-Mills fields in the diagonal gauge), as
well as $N^\Psi$ real fermionic, anticommuting first-order fields
$\Psi_1,\dots,\Psi_{N^\Psi}$,
(e.g. Dirac, Weyl, or Majorana spinors, or Rarita-Schwinger fields; of course,
complex field components should be broken into real and imaginary part).

We collect all field components into one vector
\Beq
 \Xi =(\Xi_1,\dots,\Xi_N) = (\Phi|\Psi);
\Eeq
this will be the standard coodinate on the configuration smf
\Beq
 M=\L(\cEc\V),\quad V:= \mathbb R^{N^\Phi} \oplus\Pi\mathbb R^{N^\Psi},
 \quad \cEc\V := \cEc\otimes V
\Eeq
(cf. \ref{SpaceCic} below for the definition of $\cEc$).
The model is described by the {\em Lagrange density}, which is a real, even,
entire differential power series of the form
\Beq
 \cL[\ul\Xi] = \cL\kin[\ul\Xi] + \LInt[\ul\Xi]\in
 \mathbb C\br[[(\ul\Xi_i,\partial_\mu\ul\Xi_i)_
 {i=1,\dots,N,\ \ \mu=0,\dots,d}
 ]] _{\even,\mathbb R,\opn ent }.
\Eeq
(As usual, we call a power series in even and odd variables,
$P[y|\eta]=\sum P_{\alpha\beta}y^\alpha\eta^\beta \in
 \mathbb C[[y_1,\dots,y_m|\eta_1,\allowbreak \dots,\allowbreak \eta_n]]$,
{\em entire} iff for all $R>0$ there exists $C>0$  such that
$\br|P_{\alpha\beta}| \le C R^{-|\alpha|}$
for all $\alpha,\beta$.)

Here $\cL\kin[\ul\Xi]$ is called the {\em kinetic Lagrangian}, and
$\LInt[\ul\Xi]$ is the {\em interaction term}; we now specify our
requirements onto these terms.

For the interaction term, we require that it does not contain derivatives
of first-order fields, and is at most linear in the derivatives of
second-order fields:
\Beq
 \LInt[\ul\Xi] = \LInt'(\ul\Xi) +
 \LInt^{\mu,i}(\ul\Xi) \partial_\mu\ul\Phi_i
\Eeq
where $\LInt^{\mu,i}(\ul\Xi)$ and $\LInt'(\ul\Xi)$ are
real, even, entire (non-differential) power series of lower degree
$\ge2$ and $\ge3$, respectively. We get
\Beqn StdFEq
 \frac\delta{\delta\ul\Phi_i}\LInt[\ul\Xi] =
 \frac\partial{\partial\ul\Phi_i} \LInt'(\ul\Xi)
 + \LInt^{\mu,ij}(\ul\Xi)\partial_\mu\ul\Phi_j
\Eeq
with
\Beq
 \LInt^{\mu,ij}(\ul\Xi) :=
 \frac\partial{\partial\ul\Phi_i}\LInt^{\mu,j}(\ul\Xi)
 - \frac\partial{\partial\ul\Phi_j}\LInt^{\mu,i}(\ul\Xi).
\Eeq
Turning to the kinetic Lagrangian, we assume the standard forms which lead
to the Klein-Gordon operator as kinetic operator for second-order fields,
and to some Dirac-like operator for first-order fields:
\Beqn GenFrmOfLKin
 \cL\kin[\ul\Xi] = \frac12 \left(
  - \partial_\mu\ul\Phi_i\partial^\mu\ul\Phi_i
  - m^\Phi_{ij}\ul\Phi_i\ul\Phi_j
  + \i\partial_\mu\ul\Psi_k\Gamma^\mu_{kl}\ul\Psi_l
  + \i\ul\Psi_k m^\Psi_{kl}\ul\Psi_l \right).
\Eeq
Usually, the eigenvalues of the symmetric matrix
$m^\Phi\in\mathbb R^{N^\Phi\times N^\Phi}$
are the squared masses of the degrees of freedom;  however,
in view of the Higgs field, there is a reason not to put any requirement of
positive definiteness onto $m^\Phi$. For the matrices
$m^\Psi, \Gamma^\mu\in\mathbb R^{N^\Psi\times N^\Psi}$
we require that $(\Gamma^\mu)\trp=\Gamma^\mu$ (any antisymmetric part would
produce only a total derivative) and $(m^\Phi)\trp = -m^\Phi$. Also we suppose
that for the arising "massive Dirac operator" $\Dirac$ with
\Beqn MassDirOp
 \Dirac_{ij}:= \i\left(-\Gamma^{\mu}_{ij}\partial_\mu + m^\Psi_{ij}\right)
\Eeq
there exists another first-order operator
$K=K(\partial_t,\partial_a)$ with constant coefficients such that
\Beqn KDiracKlG
 K_{ij}\Dirac_{jk}= (\KlGord+M)\delta_{ik}
\Eeq
where, as usual, $\KlGord$ is the Klein-Gordon operator, and $M$ is some
real-valued mass matrix (no positivity condition is necessary).

Given such a model, we define the {\em underlying bosonic model}
as having field content $\Phi_1,\dots,\Phi_{N^\Phi}$
and Lagrangian $\cL\bos[\ul\Phi]:=\cL[\ul\Xi]|_{\ul\Psi=0}$.

\subsection{Configuration families and solution families}\label{SpaceCic}

We take over the notations of \cite{[CAUCHY]}, \cite{[ABSEVEQ]}:
For $r\ge0$, let
\Beqn DefbVr
 \bV_r:=\{(t,x)\in\rdmm:\quad \br|x|\le r+\br|t|\},
\Eeq
and denote temporarily by $C^\infty_{\bV_r}(\rdmm)$ the closed subspace of
$C^\infty(\rdmm)$ which consists of all those elements which have support
in $\bV_r$. Set
\Beq
 \cEc = \bigcup_{r>0} C^\infty_{\bV_r}(\rdmm)
\Eeq
and equip it with the inductive limit topology. This is a strict inductive
limes of Fr\`echet spaces, and hence complete. Also, $\cD(\rdmm)$ is dense
in $\cEc$; hence the latter space is admissible in the sense
of \cite{[FUNCT]}. Moreover, one easily shows that
the subspace $\cEc$ of $C^\infty(\rdmm)$ is invariant under the standard
action of the Poincar\'e group $\mathfrak P$, and that the arising action
is continuous.

We will call $V := \mathbb R^{N^\Phi|N^\Psi}$
the {\em field target space}. Also, we set
$\dot V:= \mathbb R^{N^\Phi}$; this will be the target space
for the velocities of the second-order fields at the Cauchy hyperplane.
Thus,  setting
\Beq
 \cEc\V:=\cEc\otimes V, \qquad \cEc\CauV:=\cD(\rdm)\otimes(V\oplus\dot V),
\Eeq
the {\em smf's of Cauchy data and of configurations}, \eqref{cEcCauSmf},
\eqref{cEcSmf} are now well-defined. Analogously, we set
\Beq
 \cE\CauV := C^\infty(\rdm)\otimes (V\oplus\dot V),\quad
 \cE\V :=C^\infty(\rdmm)\otimes V.
\Eeq

Let $Z$ be an arbitrary smf. A {\em configuration family parametrized
by $Z$} (or {\em $Z$-family}, for short) is an even, real
superfunction $\Xi'$ on $Z$ with values in the locally convex space $\cE\V$:
\Beq
 \Xi'=(\Phi'|\Psi')\in\M^{\cE\V}(Z).
\Eeq
Thus, $\Xi'$ encodes a $N$-tuple $\Xi'=(\Xi'_1,\dots,\Xi'_N)$ of
$C^\infty(\rdmm)$-valued superfunctions on $Z$.

If $\Xi'$ lies in the subspace $\M^{\cEc\V}(Z)$ we call $\Xi'$ a
{\em $Z$-family of quality $\cEc$}; these are of main interest for us.

(In \cite{[CAUCHY]}, \cite{[ABSEVEQ]} we have considered more general
families which have values in Sobolev spaces and are defined over time
intervals; however, for our purposes here, they are useless.)

Now, given an smf morphism $\pi:Z'\to Z$ we can assign as in \cite{[CAUCHY]}
to every $Z$-family $\Xi'$ its {\em pullback} $\Xi":=\pi^*(\Xi')$ which
is a $Z'$-family. In fact, the process of passing from $\Xi'$ to
$\Xi"$ means in family language nothing but a {\em change of parametrization}
(cf. \cite[1.11]{[WHAT]}).

One family of quality $\cEc$ is given a priori, namely the $M$-family
\Beq
 \Xi=(\Phi|\Psi)\in\M^{\cEc\V}(M)
\Eeq
where, we recall, $M=\L(\cEc\V)$ is the smf of configurations of quality
$\cEc$, and $\Xi$ is the standard coordinate (cf. \cite[2.5, 2.6]{[WHAT]}).

$\Xi$ is in fact the {\em universal family of quality $\cEc$}:
Given an arbitrary $Z$-family $\Xi'$ of quality $\cEc$, it defines by
\cite[2.8.1]{[WHAT]} a {\em classifying morphism}
\Beq
 \Xi':Z\to M, \quad \widehat{\Xi'} = \Xi'
\Eeq
and $\Xi'$ arises from $\Xi$ just by pullback: $\Xi'={\Xi'}^*(\Xi)$.

\Brm
In the language of category theory, this means that the
cofunctor
\Beq
 \{\text{supermanifolds}\}\to\{\text{sets}\},\qquad Z\mapsto\M^{\cEc\V}(Z),
\Eeq
is represented by the object $M$ with the universal element $\Xi$.
\Erm

Fixing a $Z$-family $\Xi'$ of quality $\cEc$, the field strengthes
$\Xi'_i(t,x)=\delta_{(t,x)}\circ\Xi'_i$ for $(t,x)\in\rdmm$ are
scalar superfunctions on $Z$. More generally, we define the
{\em value at $\Xi'$} of any superfunctional $K\in\O^F(M)$ as
the pullback of $K$ along ${\Xi'}$:
\Beq
 K[\Xi']:={\Xi'}^*(K)\in\O^F(Z).
\Eeq
For instance, in case $Z$ is a point, the value $K[\Xi']$ of
an $F$-valued superfunctional $K$ at a $Z$-family $\Xi'$ is an element of
$F_{\mathbb C}$; thus, for a scalar functional $K\in\O(M)$, it is
simply a complex number (which, however, is zero for all odd
$K$, and, in particular, for the fermionic field strengthes).

\subsection{Solution families and action principle}
A {\em $Z$-family of solutions}, or {\em solution family} for short,
is a $Z$-family $\Xi'\in\M^{\cE\V}(Z)$ which satisfies
\Beqn defSolFam
 (\delta/\delta\ul\Xi_i)\cL[\Xi']=0
\Eeq
for all $i$ (the l.~.h.~s. is obviously well-defined in $\O^{\cE\V}(Z)$).
Trivially, every pullback of a solution family is a solution family.

Of course, the universal family $\Xi$ is not a family of solutions.
However, we will show in Thm. \ref{MainThm} that in the case of a
complete model, there exists a family of solutions $\Xi\sol$ of quality
$\cEc$ which is universal for this quality, i.e. every other solution
family of quality $\cEc$ will be a pullback of $\Xi\sol$.

If $Z=P$ is a point then a $Z$-family of solutions is just an element
$\phi\in\cE\V\seven = C^\infty(\rdmm)\otimes V\seven$ which solves
the field equations of the underlying bosonic model in the usual sense.

The {\em Cauchy data} of a family $\Xi'\in\M^{\cE\V}(Z)$ is the element
\Beq
 \left(\Xi'(0),\partial_0\Phi'(0)\right)\in\M^{\cE\CauV}(Z).
\Eeq
Up to now, solution families are characterized rather formally,
as making the variational derivatives of the differential polynomial $\cL$
vanish. We now look for the action principle.

Inserting the coordinate superfunctions $\Xi_i\in\O^{\cEc}(M)$ into the
Lagrangian, we get the {\em action density} $\cL[\Xi]\in\M^{\cEc}(M)$ as an
$\cEc$-valued superfunction on the configuration smf $M$; however, its
space-time integral,
\Beqn TotAction
 S := \int dx\cL[\Xi](x),
\Eeq
is ill-defined. What we can do is to define the action over any
closed subset $\Omega\seq\rdm$ which has the property that
$\Omega\cap\bV_r$ is compact for all $r>0$ (cf. \eqref{DefbVr}):
\Beq
 S_\Omega[\Xi] := \int_\Omega dx\cL[\Xi](x) \in\M(M).
\Eeq
In particular, the action within a time interval $[t_0,t_1]$ is well-defined:
\Beq
 S_{t_0,t_1}[\Xi] = \int_{t_0}^{t_1}dt\int_{\rdm} dx\cL[\Xi](t,x)\in\M(M).
\Eeq
Note that on the smf of smooth configurations $M\scinf = \L(\cE\V)$
(cf. \ref{CInfSolSmf} below), the action $S_\Omega[\Xi]\in\M(M\scinf)$
is defined only if $\Omega$ is compact; in particular, $S_{t_0,t_1}[\Xi]$
is not well-defined as superfunction on $M\scinf$.

For a superfunction with values in continuous functions on $\rdmm$,
i.e. $K\in\O^{C(\rdmm)}(Z)$, let, as in \cite{[CAUCHY]},
the {\em target support of $K$} be defined as
\Beq
 \opn t-supp K := \opn Closure \Bigl(\{x\in\rdmm:\ \ K(x) \not= 0\}\Bigr),
\Eeq
where, of course, $K(x) = \delta_x\circ K$. This should not be confused
with the support of a power series as defined in \FUNref{3.11}.

Now $\Xi'$ is a solution family iff $S_\Omega[\Xi']$ remains
stationary with respect to infinitesimal increments of $\Xi'$ which have
their target support in the interior $\Omega^0$:

\begin{prp}
For a $Z$-family of configurations $\Xi'\in\M^{\cE\V}(Z)$
the following conditions are equivalent:

(i) $\Xi'$ is a $Z$-family of solutions;

(ii) We have $(\partial_\xi S_\Omega)[\Xi'] = 0$
for all $\Omega$ as above and $\xi\in\cD(\rdmm)\otimes V$ with
$\supp \xi\seq\Omega^0$. Here $\partial_\xi$ is the directional
derivative, cf. \FUNref{3.9, 3.12};

(iii)
We have $\chi(S_\Omega)[\Xi'] = 0$ for all $\Omega$ as above and
$\chi\in\cX_{\cEc\V}((\cEc\V)\seven)$
with $\opn t-supp \chi(\Xi_i)\seq\Omega^0$ for all $i$.
Here $\cX_{\cEc\V}((\cEc\V)\seven)$
denotes the space of all infinitesimal
transformations over $(\cEc\V)\seven$, cf. \FUNref{3.12}.
\end{prp}

\begin{proof}
We first need a formula for $\chi(S_\Omega)[\Xi']$ where
$\chi$ is as in (iii): Mimicking the proof of \FUNref{Thm. 2.8}, we have

\Beqn LambdAct
 \chi(S_\Omega)[\Xi'] =\sum_i \int_\Omega  dx\
 \left(
  \chi(\Xi_i)[\Xi'](x)\, \frac\delta{\delta\ul\Xi_i}\cL[\Xi'](x)
  +  \partial_\mu \left( \chi(\Xi_i)[\Xi'](x)\,
  \frac\partial{\partial(\partial_\mu\ul\Xi_i)}\cL[\Xi'](x) \right)
 \right).
\Eeq
Now if $\Xi'$ is a solution family then the first term of
the sum on the r.~h.~s. vanishes by \eqref{defSolFam}
while the second one is a total differential with target support
in the interior of $\Omega$; hence its integral vanishes, too,
proving (i)$\Rightarrow$(iii).

(iii)$\Rightarrow$(ii) is obvious.

To show (ii)$\Rightarrow$(i), we specialize \eqref{LambdAct} to
$\chi=\partial_\xi$, getting
$0=\sum_i \int_\Omega  dx\ \xi_i(x)\frac\delta{\delta\ul\Xi_i}\cL[\Xi'](x)$
for all $\Omega,\xi$ as in (ii); this implies \eqref{defSolFam}.
\end{proof}

\Brm
One can give \eqref{TotAction} sense as scalar superfunction
$S\in\M(M_{C^\infty_0})$ on the smaller configuration smf
$M_{C^\infty_0} :=\L(C^\infty_0(\rdmm)\otimes V)$ (which, of course, is
too small to contain non-trivial solutions of the field equations).
Taking its exterior differential, alias total variation (cf. \FUNref{3.8}),
\Beq
 \delta S = \delta S[\Xi,\delta\Xi] = \sum_{i=1}^N
  \int_{\rdmm} dx\; \delta\Xi_i(x)\,\frac\delta{\delta\Xi_i(x)}S,
\Eeq
and using the notations of \cite[5.5, 6.6]{[IS]}, we have
\Beq
  \delta S\in\Omega^1(M_{C^\infty_0})
  \ = \ \sideset{^{C^\infty_0(\rdmm)\otimes V}}{^{\Bbb R}}\M (M_{C^\infty_0}).
\Eeq
It is easy to see that, with respect to the embedding of smf's
$M_{C^\infty_0} \too\seq M \too\seq M\scinf$,
this lifts to unique elements
\Beq
\delta S \in
 {\phantom. \sideset{^{C^\infty_0(\rdmm)\otimes V}}{^{\Bbb R}}\M (M)},\quad
\delta S \in
 {\phantom. \sideset{^{C^\infty_0(\rdmm)\otimes V}}{^{\Bbb R}}\M (M\scinf)}.
\Eeq
Note, however, that $\delta S$ does not lie in the subspace
\phantom.$\sideset{^{\cEc}}{^{\Bbb R}}\M (M) = \Omega^1(M)$, i. e.,
it is not a one form on $M$ (and the less on $M\scinf$), and thus
there is no contradiction to the Poincar\'e Lemma given in \cite[6.9]{[IS]}.

Now a $Z$-family of configurations $\Xi'\in\M^{\cE\V}(Z)$
is a $Z$-family of solutions iff the pullback
${\Xi'}^*(\delta S) = \delta S[\Xi',\delta\Xi] \in
 \phantom.\sideset{^{C^\infty_0(\rdmm)\otimes V}}{^{\Bbb R}}\M (Z)
$
vanishes.
\Erm

\subsection{Reduction to first-order equations}

The relevant variational derivatives of the Lagrangian take the form
\Bal
 \frac\delta{\delta\ul\Phi_i}\cL[\Xi]&=
 \KlGord\Phi_i - m^\Phi_{ij}\Phi_j
  + \frac\partial{\partial\ul\Phi_i} \LInt'(\Xi)
  + \LInt^{\mu,ij}(\Xi)\partial_\mu\Phi_j,
 && (i=1,\dots,N^\Phi)
 \\
 \frac\delta{\delta\ul\Psi_i}\cL[\Xi]&=
 \Dirac_{ij}\Psi_j + \frac\partial{\partial\cj\Psi_i}\LInt[\Xi]
 && (i=1,\dots,N^\Psi)
\Eal
as identities in $\O^{\cEc}(M)$. In a standard way, we write the field
equations as first-order evolution system:
\begin{subequations}
\Baln 1stOrdSystem
 \partial_t\Phi_i &=\dot\Phi_i,
 \\
 \partial_t\dot\Phi_i&=
 \Delta\Phi_i - m^\Phi_{ij}\Phi_j
  + \frac\partial{\partial\ul\Phi_i} \LInt'(\Xi)
  + \LInt^{\mu,ij}(\Xi)\partial_\mu\Phi_j,
 \\\label{EvPsi}
 \partial_t\Psi_i &= -{(\Gamma^0)^{-1}}_{ij}
 \Bigl( \Gamma^a_{jk}\partial_a \Psi_k  + m^\Psi_{jk} \Psi_k
 + \i\frac\partial{\partial\ul\Psi_j}\LInt[\Xi] \Bigr)
\Eal
\end{subequations}
(for legibility, we have suppressed the  apostrophes which indicate that
these are conditions for a configuration family).
Thus, $\Xi'$ is a solution family of our model iff $(\Xi',\partial_t\Phi')$
is a  solution family of the system \eqref{1stOrdSystem}--\eqref{EvPsi}.

We get a system of field equations for the first-order field components
\Beq
 (\Xi_1\fo,\dots,\Xi_N\fo)
 = (\Phi_i,\dot\Phi_i|\Psi_j),
 \quad N\fo:=2N^\Phi+N^\Psi.
\Eeq

\begin{lem}\label{ReductionLem}
The system \eqref{1stOrdSystem}--\eqref{EvPsi} belongs to the class of systems
of
field equations considered in \cite{[CAUCHY]}, \cite{[ABSEVEQ]}, and
is causal.
\end{lem}

\begin{proof}
First we have to look for the spatially Fourier-transformed influence function
$\hat A=(\hat A_{ij})=\hat A(t,p)$, i.e. the solution of the linearized
and spatially Fourier-transformed equations \eqref{1stOrdSystem}--\eqref{EvPsi}
with the initial data
\Beq
 \hat A(0,p) = (2\pi)^{-d/2}1_{N\times N}.
\Eeq
For notational convenience, we work with the vector
$(\Phi_i,\dot\Phi_i|\Psi_j)$
of real and complex field components. Thus,
$\hat A=\opn diag ({\hat A}^\Phi,\ {\hat A}^\Psi)$ where
\Beqn IAPhi
 {\hat A}^\Phi(t,p)=
 \begin{pmatrix}
  \partial_t\hat\cA^\Phi & \hat\cA^\Phi \\
  {\partial_t}^2\hat\cA^\Phi  & \partial_t\hat\cA^\Phi
 \end{pmatrix},
\Eeq
and $\cA^\Phi$ is the solution of
\Beq
 (\partial_t^2 + p^2 +m^\Phi)\hat\cA^\Phi(t,p)=0, \quad
 \hat\cA^\Phi(0,p)=(2\pi)^{-d/2}1_{N^\Phi\times N^\Phi}.
\Eeq
Explicitly,
\Beq
 \hat\cA^\Phi(t,p) = (2\pi)^{-d/2} f_t(p^2+m^\Phi),\qquad
 f_t(z):=\frac{\sin t\sqrt z}{\sqrt z}.
\Eeq
Thus, if $m^\Phi=\opn diag (\mu_1,\dots,\mu_{N^\Phi})$ then the inverse
spatial Fourier transform of $\hat\cA^\Phi$ becomes
$\cA^\Phi(t,x) = \opn diag (D_{\mu_1}(t,x),\allowbreak\dots,\allowbreak
 D_{\mu_{N^\Phi}}(t,x))
$
where $D_{\mu_i}(t,x)$ is the {\em Pauli-Jordan exchange function} with
mass $\mu_i$.

In the first-order sector, if writing $\Dirac=\Dirac(\partial_t,\partial_x)$
then
${\hat A}^\Psi$ is determined by
\Beq
 \Dirac(\partial_t,\i p){\hat A}^\Psi(t,p)=0,\quad
 {\hat A}^\Psi(0,p)=(2\pi)^{-d/2}1_{N^\Psi\times N^\Psi}.
\Eeq
We claim that the solution is explicitly given by
\Beqn IAPsi
 {\hat A}^\Psi(t,p):= (2\pi)^{-d/2}
 K(\partial_t,\i p_a) \cos t\sqrt{p^2+m^\Psi} \cdot\Dirac(0,\i p_a)/(p^2+M)
\Eeq
where $K(\partial_t,\partial_a)$ is the operator from \eqref{KDiracKlG}.
Indeed,
\Beq
\begin{split}
{\hat A}^\Psi(0,p)
 &= (2\pi)^{-d/2} K(\partial_t,\i p_a)(1_{N^\Psi\times N^\Psi})
  \Dirac(0,\i p_a)/(p^2+M)\\
 &=(2\pi)^{-d/2} K(0,\i p_a)\Dirac(0,\i p_a)/(p^2+M) 1_{N^\Psi\times N^\Psi}
  =(2\pi)^{-d/2} 1_{N^\Psi\times N^\Psi}.
\end{split}
\Eeq
Having constructed $\hat A$, one now verifies the estimates
\begin{subequations}
\Baln TheCAEst0
 \br\| \hat A^\Phi(t,p) \|
   &\le \frac K{1+|p|},
\\
 \br\| \frac d{dt} \hat A^\Phi(t,p) \|
  &\le K, \label{TheCAEst1}
\\
 \br\| \frac{d^2}{dt^2} \hat A^\Phi(t,p) \|
  &\le K(1+|p|), \label{TheCAEst2}
\\
 \br\| \hat A^\Psi(t,p) \| &\le K \label{ThefAEst}
\Eal
\end{subequations}
valid for (say) $t\in(-1,1)$ and suitable $K>0$.
This allows to take the vector of smoothness offsets $\tau$
which assigns $1$ to the second-order
field components $\Phi_i$, and $0$ to their velocities  $\dot\Phi_i$
as well as to the first-order fields $\Psi_j$.

Also, using the Paley-Wiener Theorem, we have for the inverse Fourier
transform $A$ of $\hat A$
\Beq
  \supp A \seq \{(t,x)\in\rdmm:\quad \br|x|\le \br|t|\}.
\Eeq
so that our system is causal.
\end{proof}

\Brm
In view of the expectation that particles with imaginary rest
mass should move tachyonically, it is somewhat astonishing that
the scalar fields $\Phi_i$ have causal propagation even if $m^\Phi$ has
negative eigenvalues. Of course, the corresponding components (after
diagonalization) cannot describe a physically senseful free field;
attempting to canonically quantize it would lead to a state space with
indefinite metric.  (Complex eigenvalues cannot occur due to symmetry of
$m^\Phi$.)
\Erm

\subsection{The formal solution}
In the following, we look at the solution of the Cauchy problem for the
field equations on the formal power series level. Essentially, one only
needs to repeat the approach of \cite{[ABSEVEQ]}, \cite{[CAUCHY]}, with
adapting it to our context which contains first as well as second-order
fields. The result is:

\begin{cor}
There exists a uniquely determined formal power series (in the sense of
\FUNref{2.3}),
\Beq
 \Xi\sol = \Xi\sol[\Xi\Cau,\dot\Phi\Cau]\in{\mathcal
P}_f(\cEc\CauV;\cEc\V)\sevR
\Eeq
which solves the Cauchy problem \eqref{XiSolIsSol}, \eqref{XiSolRightCD}.
Letting $\Xi\sol=\sum_{m>0}\Xi\sol_{(m)}$ denote its splitting into
homogeneous components, the first one, $\Xi\sol_{(1)}$, is the formal
solution of the corresponding free model:
\Beq
 \begin{split}
 &(\Phi\sol_{(1)})_i[\Xi\Cau,\dot\Phi\Cau](t,y)
  = \int dx \left(\partial_t \cA^\Phi_{ij}(t,y-x)\Phi\Cau_j(x)
  + \cA^\Phi_{ij}(t,y-x)\dot\Phi\Cau_j(x)\right),
 \\
 &(\Psi\sol_{(1)})_i[\Xi\Cau,\dot\Phi\Cau](t,y)
 = \int dx A^\Psi_{ij}(t,y-x)\Psi\Cau_j(x),
 \end{split}
\Eeq
while the higher components are recursively given by
\Beq
 (\Xi\sol_{(m)})_i(t,y)=
 \int_{\mathbb R\times\rdm}
 dsdx G_{ij}(t,s,y-x)
 \frac\delta{\delta\ul\Xi_j}\LInt[\Xi\sol_{(<m)}]_{(m)}(s,x)
\Eeq
for $m\ge2$ where, of course, $\Xi\sol_{(<m)}=\sum_{n=1}^{m-1}\Xi\sol_{(n)}$,
and
\Beq
 G(t,s,x):= \bigl(\theta(t-s)-\theta(s)\bigr)\opn diag
 \Bigl(\cA^\Phi(t-s,x),\ A^\Psi(t-s,x)\cdot(\Gamma^0)^{-1}\Bigr)
\Eeq
where $\theta(s)$ is $1,\ 1/2,\ 0$ for $t>0$, \ $t=0$, \ $t<0$, respectively.
\qed
\end{cor}

\Brm
The Green function matrix $G:=(G_{ij})= \opn diag (G^\Phi,\ G^\Psi)$
is characterized by
\Beq
 \opn diag (\KlGord,\ \Dirac)_{t,x} G(t,s,x) =
 -\delta(t-s)\delta(x)\cdot 1_{N^\Phi+N^\Psi},\qquad
 G(0,s,x)=0, \quad \partial_tG^\Phi(0,s,x)=0.
\Eeq
\Erm

The formal power series $\Xi\sol$, which we call the {\em formal solution
of the classical model}, will be the power series expansion at
the zero configuration of the solution of the "analytic Cauchy problem"
\eqref{AnCauProbl}, \eqref{XiSolIsSol}, \eqref{XiSolRightCD}.

\subsection{Cauchy uniqueness and causality}\label{CauUnCaus}

The first thing to observe is that solution families are uniquely determined
by their Cauchy data. Before giving a "localized" variant of this fact which
also expresses the causality of the propagation of perturbations, we need
some notations:

For $(s,x),(t,y)\in\rdmm$ we will write $(s,x)\causall(t,y)$ iff $(t,y)$ lies
in the forward light cone of $(s,x)$, i.~e. $(t-s)^2\ge(y-x)^2$.

Given a point $p=(s,x)\in\rdmm$ with $s\not=0$,  write
\Bal
 &\Omega(p) :=
 \begin{cases}
  \{(s',x')\in\rdmm:\ (s',x') \causall (s,x),\ 0 < s'\}  & \text{if $s>0$,}\\
  \{(s',x')\in\rdmm:\ (s,x) \causall (s',x'),\ s' < 0\}  & \text{if $s<0$,}
 \end{cases}
 \\
 &\J(p) := \{x'\in\rdm:\ \br|x'-x| < \br|s|\}.
\Eal
By the results of \cite{[CAUCHY]}, \cite{[ABSEVEQ]}, causality implies
that perturbations of solution families propagate within the light cone,
as to be expected:

\begin{thm}\label{CausCauUniqu}
Let be given a point $p=(s,x)\in\rdmm$ with $s\not=0$, and
let be given two $Z$-families $\Xi',\Xi"\in\M^{\cE\V}(Z)$. Suppose that
\Bal
 &\frac\delta{\delta\ul\Xi_i}\cL[\Xi']|_{\Omega(p)}
 = \frac\delta{\delta\ul\Xi_i}\cL [\Xi"]|_{\Omega(p)} =0
 \quad (i=1,\dots, N),
 \\
 &\left((\Xi'-\Xi")(0),\partial_0(\Psi'-\Psi")(0)\right)
  |_{\J(p)}=0.
\Eal
Then $(\Xi'-\Xi")|_{\Omega(p)}=0$.
\qed\end{thm}

\subsection{Completeness, universal solution family, and the smf of
classical solutions}
Loosely said, we call the model complete iff the underlying bosonic
model is globally solvable:

\begin{thm}\label{ComplThm}
The following conditions are equivalent:

(i) For every smooth solution
$\phi\in C^\infty((a,b)\times \mathbb R^d) \otimes V\seven$ of
the underlying bosonic field equations
\Beqn UnderlBFE
 \frac\delta{\delta\ul\Phi_i(x)} \cL\bos[\phi]=0
\Eeq
on a bounded open time interval $(a,b)$ such that $\supp\phi(t)$ is
compact for some (and hence all) all $t\in(a,b)$,
there exists a Sobolev index $k>d/2$ such that
\Beqn APEst
 \sup_{t\in(a,b)} \max\br\{\|\phi_i(t)\|_{H_{k+1}(\rdm)},\
 \|\partial_t\phi_i(t)\|_{H_{k}(\rdm)}\} < \infty
\Eeq
for all $i=1,\dots,N^\Phi$.

(ii) The underlying bosonic equations are all-time solvable with
quality $\cEc$:

Given bosonic Cauchy data
$(\phi\Cau,\dot\phi\Cau)\allowbreak \in(\cEc\CauV)\seven$
there exists an element $\phi\in(\cEc\V)\seven$
with these Cauchy data which solves \eqref{UnderlBFE}.

\noindent If these conditions are satisfied we call the model {\em complete}.
\end{thm}

\Brmn NoDerivCpl
In the absence of derivative couplings, $\LInt^{\mu,i}=0$, it is
sufficient to have \eqref{APEst} for some $k>d/2-1$.
\Erm

\begin{proof}
In view of Lemma \ref{ReductionLem}, the Theorem immediately follows from the
corresponding results of  \cite{[CAUCHY]}, \cite{[ABSEVEQ]}.
\end{proof}

For more comment, cf. \cite{[CAUCHY]}.

The central result of this section is:

\begin{thm}\label{MainThm}
Suppose that the model is complete.

(i) The formal solution $\Xi\sol$ is the Taylor expansion at zero of a
unique superfunctional
\Beqn SFuncXiSol
 \Xi\sol[\Xi\Cau,\dot\Phi\Cau]\in\M^{\cEc\V}(M\Cau),
\Eeq
and \eqref{SFuncXiSol} is an $M\Cau$-family of solutions of quality $\cEc$.

(ii) The image of the arising smf morphism (cf. \cite[2.8,2.12]{[WHAT]})
$\Xi\sol: M\Cau\to M$ is a split sub-smf which we call the
{\em smf of classical solutions}, or, more exactly, the
{\em smf of smooth classical solutions with spatially compact support},
and denote by $M\sol\seq M$.

(iii) $M\sol$ has the following universal property:
Recall that fixing an smf $Z$ we have a bijection between $Z$-families
$\Xi'$ of configurations of quality $\cEc$, and morphisms $\Xi':Z\to M$.
Now $\Xi'$ is a solution family iff $\Xi'$ factors to $\Xi':Z\to M\sol\seq M$.

In this way, we get a bijection between $Z$-families
$\Xi'$ of solutions of quality $\cEc$ and morphisms $\Xi':Z\to M\sol$.
\end{thm}

\begin{proof}
In view of Lemma \ref{ReductionLem}, this follows from the corresponding
results of  \cite{[CAUCHY]}, \cite{[ABSEVEQ]}.
\end{proof}

\Brm
(1) In the language of category theory, assertion (v) means that the
cofunctor
\Beq
 \{\text{supermanifolds}\}\to\{\text{sets}\},\qquad
 Z\mapsto\{\text{supermanifolds}\}
\Eeq
is represented by the object $M$ with the universal element $\Xi\sol$. Thus,
\eqref{SFuncXiSol} is the {\em universal family of solutions of quality
$\cEc$}: Every other family of solutions of this quality arises
uniquely as pullback from \eqref{SFuncXiSol}.

(2) Taking in (iii) $P$-families where $P$ is a single point we get that
the underlying manifold $\widetilde{M\sol}$ is just the set of all
bosonic configurations $\phi\in (\cEc\V)\seven$
which satisfy the underlying bosonic field equations
\eqref{UnderlBFE}. The underlying map of $\Xi\sol: M\Cau\to M$ assigns to each
bosonic Cauchy datum $(\phi\Cau,\dot\phi\Cau)$ the unique solution
$\phi$ of \eqref{UnderlBFE}
with $\phi(0)=\phi\Cau$,\ \ $\partial_t\phi(0) = \dot\phi\Cau$.

(3) Note that $M\sol$ is still a linear smf which is, however, in a
non-linear way embedded into $M$. In a forthcoming paper we will show that,
once the action of the Lorentz group on $V$ has been fixed, the
sub-smf $M\sol$ is invariant under the arising action of the Poincar\'e group
on $M$; the other data $\alpha$, $\Xi\free$, $\Xi\sol$ are not
(they are only invariant under the Euclidian group of $\rdm$).

(4) The superfunction \eqref{SFuncXiSol} is uniquely characterized by the
conditions \eqref{XiSolIsSol}, \eqref{XiSolRightCD}.
The latter property can be recoded supergeometrically to the fact that the
composite morphism
\Beq
  M\Cau \too{\Xi\sol} M \too\pi M\Cau
\Eeq
is the identity; here $\pi$ is the projection onto the Cauchy data:
$\hat\pi[\Xi] = (\Xi(0),\partial_0\Phi(0))$.

(5) In qft slang, the homomorphism
\Beqn RestMSh
 \O^F(M)\to\O^F(M\sol),\qquad K[\Xi]\mapsto K[\Xi\sol]
\Eeq
is called {\em restriction of classical observables onto the mass shell}
(the latter term comes from free field theory). It follows from ass. (iii)
that \eqref{RestMSh} is surjective.

(6) Let us comment on the fact that completeness depends only on the
underlying bosonic model: Mathematically, this is an analogon of several
theorems in supergeometry that differential-geometric
tasks, like trivializing a fibre bundle, or presenting a closed form
as differential, are solvable iff the underlying smooth tasks are solvable.

Physically, our interpretation is somewhat speculative:
In the bosonic sector, the classical field theory approximates the behaviour
of coherent states, and completeness excludes that "too many" particles may
eventually assemble at a space-time point, making the state non-normable.
On the fermionic side, apart from the non-existence of genuine coherent
states, it is the Pauli principle which automatically prevents such an
assembly.
\Erm

\subsection{The smf of classical solutions without support restriction}
\label{CInfSolSmf}

It takes not much additional effort to lift the constraints on the
supports of solution families, considering arbitrary smooth solution families.
The appropriate smf's of Cauchy data and configurations are
\Beq
 M\Cau\scinf := \L(\cE\CauV),\quad M\scinf := \L(\cE\V).
\Eeq
In view of Lemma \ref{ReductionLem}, the results of \cite{[CAUCHY]},
\cite{[ABSEVEQ]} yield:

\begin{thm}\label{MainThmSm}
Suppose the model is complete. Then $\Xi\sol$ extends uniquely to a
superfunctional
\Beq
 \Xi\sol\in\M^{\cE\V}(M\CauV\scinf).
\Eeq
Moreover, the image of the arising smf morphism
\Beq
 \Xi\sol: M\Cau\scinf = \L(\cE\CauV)\to\L(\cE\V) = M\scinf
\Eeq
is a split sub-smf which we call the
{\em smf of smooth classical solutions (without support restriction)},
and denote by $M\sol\scinf\seq M\scinf$.
\qed\end{thm}

Also, the remaining properties of $M\sol\scinf$ are completely
analogous to that of $M\sol$.
We get a commutative diagram of smf's
\Bcd
 M\Cau           @>\Xi\sol>> M\sol           @>\seq>> M\\
 @V\seq VV                         @V\seq VV                @V\seq VV\\
 M\Cau\scinf@>\Xi\sol>> M\sol\scinf@>\seq>> M\scinf.
\Ecd

\subsection{Local excitations}\label{LocExc}
A further variant arises by considering compactly supported excitations of
a classical solution; in particular, it is applicable for situations with
spontaneous symmetry breaking, like the Higgs mechanism.
In view of Lemma \ref{ReductionLem}, the results of \cite{[CAUCHY]}
yield:

\begin{thm}\label{MainThmLocEx}
Suppose that the model is complete, and fix a
solution $\phi\in {\cE\V}\seven$ of the underlying bosonic field equations
\eqref{UnderlBFE}; let $(\phi\Cau,\dot\phi\Cau)$ be its Cauchy data.

(i) The superfunctional
\Beqn DefXiExc
 \Xi\exc_\phi[\Xi\Cau,\dot\Phi\Cau] :=
 \Xi\sol[\Xi\Cau+\phi\Cau,\dot\Phi\Cau+\dot\phi\Cau]-\phi,
\Eeq
which lies a priori in $\M^{\cE\V}(M\Cau\scinf)$, restricts to a
superfunctional
\Beq
 \Xi\exc_\phi[\Xi\Cau,\dot\Phi\Cau]\in\M^{\cEc\V}(M\Cau).
\Eeq

(ii) The image of the arising smf morphism $\Xi\exc_\phi: M\Cau \to M$
is a split sub-smf which we call the {\em smf of excitations around $\phi$},
and denote by $M\exc_\phi\seq M$.

(iii) $M\exc_\phi$ has the following universal property:
Given a $Z$-family $\Xi'\in\M^{\cEc\V}(Z)$, the corresponding morphism
$\Xi':Z\to M$ factors through $M\exc_\phi$ iff
the $Z$-family $\Xi'+\phi\in\M^{\cE\V}(Z)$ is a solution family.
\qed\end{thm}

\Brm
This theorem yields new information only if the $(\phi\Cau,\dot\phi\Cau)$
are not compactly carried. If they are, i.~e.
$(\phi\Cau,\dot\phi\Cau)\in(\cEc\CauV)\seven$,
then \eqref{DefXiExc} is already a priori defined as element of
$\M^{\cEc\V}(M\Cau)$, and $M\exc_\phi$ can be identified with $M\sol$.
\Erm

\subsection{Simple sufficient completeness criteria}

We call a measurable function $W:\mathbb R^m\to \mathbb R$ {\em of
polynomial growth} if it obeys an estimate
\Beqn  PolGrwth
 \br|W(y)| \le C(1 + \br\|y\|^N)
\Eeq
with some $C,N>0$.

\begin{prp}\label{SimpSuffCplCrit}
Suppose that for every solution
$\phi\in C^\infty((a,b)\times \mathbb R^d) \otimes V\seven$
of the underlying bosonic field equations \eqref{UnderlBFE} for which
$\supp\phi(t)$ is compact for all $t\in(a,b)$, we have
\Beqn L2Bounded
 \sup_{t\in(a,b)}
 \left(\br\|\partial_t\phi_i(t)\|_{L_2(\rdm)}
 +\sum_{a=1}^d \br\|\partial_a\phi_i(t)\|_{L_2(\rdm)}\right) < \infty
\Eeq
for $i=1,\dots,N^\Phi$. Also, suppose that either

(i) $d\le 1$, or

(ii) $d=2$, the interaction of the underlying
bosonic model has the form $\LInt[\ul\Phi|0]=V(\ul\Phi)$ with an
entire power series $V(\ul\Phi)$ (thus, derivative couplings are not allowed),
and the functions $\partial_{\ul\Phi_i}V(\ul\Phi)$,
$\partial_{\ul\Phi_i}\partial_{\ul\Phi_j}V(\ul\Phi)$
are all of polynomial growth, or

(iii) $d=3$, and the interaction of the underlying
bosonic model has the form
\Beq
 \LInt[\ul\Phi|0] = \sum_{i,j,k} c_{ijk}\ul\Phi_i\ul\Phi_j\ul\Phi_k
 + \sum_{i,j,k,l} c_{ijkl}\ul\Phi_i\ul\Phi_j\ul\Phi_k\ul\Phi_l
\Eeq
with real numbers $c_{ijk},c_{ijkl}$.

Then the model is complete.
\end{prp}

\Brm
(1) Thus, in $d=3$, couplings of at most fourth order are allowed. Remarkably,
this is at the same time the condition for power-counting renormalizability.
(2)
The hypothesis \eqref{L2Bounded} is usually guaranteed by energy conservation.
\Erm

\begin{proof}
For notational convenience, we assume from the beginning that fermionic
fields are absent. As in \cite{[CAUCHY]}, \cite{[ABSEVEQ]}, we use the
Sobolev norm
\Beq
 \br\|(\phi\Cau,\dot\phi\Cau)\|_{\cH\CauV_{k}} :=
 \sum_{i=1}^{N^\Phi}  \br(\br\|\phi\Cau_i\|_{H_{k+1}}
 + \br\|\dot\phi\Cau_i\|_{H_{k}} ).
\Eeq
In view of  Thm. \ref{ComplThm} and Remark \ref{NoDerivCpl}, it is in all
three cases sufficient to show
\Beqn CplCritIndHyp
 \sup_{t\in(a,b)} \br\|(\phi(t),\partial_t\phi(t))\|_{\cH\CauV_k} < \infty.
\Eeq
for $k=1$.

First we show \eqref{CplCritIndHyp} for $k=0$:  We have
$\partial_t\br\|\phi_i(t)\|_{L_2(\rdm)} \le
 \br\|\partial_t\phi_i(t)\|_{L_2(\rdm)} < C
$
for all $i$ with some $C>0$, and hence
\Beq
 \br\|\phi_i(t)\|_{L_2(\rdm)} \le\br\|\phi_i((a+b)/2)\|_{L_2(\rdm)}
 +C \br|t- (a+b)/2|
\Eeq
is bounded, which implies the assertion.

For the step to $k=1$, we apply  \cite[Lemma 4.2.3]{[ABSEVEQ]}
with respect to the norms
$\br\|\cdot\|_{\cH\CauV_1}\le\br\|\cdot\|_{\cH\CauV_0}$ within
(say) the Banach space $(\cH\CauV_2)\seven$ which is the completion of
$(\cEc\CauV)\seven$ with respect to $\br\|\cdot\|_{\cH\CauV_{2}}$.
Thus, it is sufficient to show that there exists a monotonously increasing
function $F:\mathbb R_+\to \mathbb R_+$ with
\Beqn SlfSmEst
\br\|\frac\delta{\delta\ul\Phi_i}\LInt[\phi](t)\|_{H_1}
 \le
 \bigl(1+\br\|(\phi(t),\partial_t\phi(t))\|_{\cH\CauV_{1}}\bigr)
 F(\br\|(\phi(t),\partial_t\phi(t))\|_{\cH\CauV_{0}})
\Eeq
for all $t\in(a,b)$,\ \ $i=1,\dots,N^\Phi$.

Ad (i). In $d\le 1$, we have a continuous embedding
$H_1(\rdm)\seq L_\infty(\rdm)$.
Obviously, pointwise multiplication makes $L_\infty(\rdm)$ a Banach algebra,
and yields a continuous pairing
$L_2(\rdm)\times L_\infty(\rdm) \to L_2(\rdm)$.

{}From \eqref{StdFEq} we have for $a=1,\dots,d$
\Beqn PtlAStdCpl
 \partial_a \frac\delta{\delta\ul\Phi_i}\LInt[\phi](t) =
 \partial_a\phi_l(t)
 \left(
  \frac{\partial^2}{\partial\ul\Phi_i\partial\ul\Phi_l} \LInt'(\phi(t))
  + \frac\partial{\partial\ul\Phi_l}\LInt^{\mu,ij}(\phi(t))
  \partial_\mu \phi_j(t)
 \right)
 + \LInt^{\mu,ij}(\phi(t)) \partial_a\partial_\mu\phi_j(t)
\Eeq
(note that the sum over $\mu$ ranges $0,\dots,d$). For $s>0$, let
\Bml
 F'(s):= \sup\nolimits_{(\xi,\dot\xi)\in\cH_0\CauV;\ \ \br\|(\xi,\dot\xi)\|\le
s}
  \quad \max\nolimits_{\mu,i,l} \quad \max
 \\
 \br\{
 \br\|
  \frac{\partial^2}{\partial\ul\Phi_i\partial\ul\Phi_l} \LInt'(\xi)
 \|_{L_\infty},
  \br\| \frac\partial{\partial\ul\Phi_l}\LInt^{c,ij}(\xi) \partial_c \xi_j
   + \frac\partial{\partial\ul\Phi_l}\LInt^{0,ij}(\xi) \dot\xi_j
 \|_{L_2},
 \br\|\LInt^{\mu,il}(\xi)\|_{L_\infty}
 \}.
\Eml
By our remarks above, this is finite. From \eqref{PtlAStdCpl} we have
\Bal
 \br\|\partial_a \frac\delta{\delta\ul\Phi_i}\LInt[\phi](t)\|_{L_2}
 &\le
 C F'(\br\|(\phi(t),\partial_t\phi(t))\|_{\cH\CauV_0})
 \left(\sum_a \br\|\partial_a \phi_k(t) \|_{L_2}
 + \sum_{a,\mu,j} \br\|\partial_a\partial_\mu \phi_j(t) \|_{L_2}
 \right)
 \\
 &\le
 C F'(\br\|(\phi(t),\partial_t\phi(t))\|_{\cH\CauV_0})
 \br\|(\phi(t),\partial_t\phi(t))\|_{\cH\CauV_1},
\Eal
yielding the estimate \eqref{SlfSmEst} wanted.

Ad (ii). Recall that we have a continuous embedding
\Beqn LH1Est
 H_1(\mathbb R^2)\seq L_p(\mathbb R^2)
\Eeq
for all $p\ge2$.
Also, we note that if $W$ is of polynomial growth and satisfies
$W(0)=0$, the estimate \eqref{PolGrwth}  implies that for all $p>1$
there exist $C_1,C_2$ such that

\Beq
 \br\|W(\xi_1,\dots,\xi_m)\|^p_{L_p} \le C_1 +
  C_2\sum_i\br\|\xi_i\|^{Np}_{L_{Np}}.
\Eeq
for $\xi_1,\dots,\xi_m\in H_1$. Combining this with \eqref{LH1Est} we
get that there exist $C_3,C_4$ such that
\Beqn WLH1Est
 \br\|W(\xi_1,\dots,\xi_m)\|_{L_p} \le C_3 + C_4\sum_i\br\|\xi_i\|^N_{H_1}.
\Eeq
Applying \eqref{LH1Est} and \eqref{WLH1Est}, we get an estimate
\Beq
 \br\|\partial_a(\partial_iV(\phi(t)))\|_{L_2}
 \le
  \sum_j \br\|\partial_a\phi_j(t)\|_{L_4}
  \br\|\partial_i\partial_jV(\phi(t))\|_{L_4}
 \le
  \sum_j\br\|\phi_j(t)\|_{H_2} \bigl(C_5
   + C_6\sum_k \br\|\phi_k(t)\|^N_{H_1}\bigr).
\Eeq
Applying also \eqref{WLH1Est} with $W:=\partial_i V$,
we get the estimate \eqref{SlfSmEst} needed:
\Beq
 \br\|\partial_iV(\phi(t))\|_{H_1} \le
 \bigl(1+ \sum_j\br\|\phi_j(t)\|_{H_2})
    (C_7 + C_8\sum_j\br\|\phi_j(t)\|^N_{H_1}\bigr).
\Eeq

Ad (iii).
Choose $R>0$ such that $\supp\phi(t)\seq\{x:\ \br\|x\|<R\}$ for all
$t\in(a,b)$.
It is a standard fact that we have a continuous embedding
\Beq
  H_{1}(\mathbb R^3)\seq L_6(\mathbb R^3),
\Eeq
Also,  for each $p\le 6$ we have a  continuous embedding
\Beq
  \bigl\{\xi\in L_6(\mathbb R^3):\ \  \supp\xi\seq\{x:\ \br\|x\|<R\}\bigr\}
  \seq L_p(\mathbb R^3),
\Eeq
Finally, we recall that pointwise multiplication yields a continuous bilinear
map
\Beq
 L_p(\mathbb R^3)\times L_q(\mathbb R^3)\to L_{1/(1/p+1/q)}(\mathbb R^3).
\Eeq
It follows that
\Beq
 \br\|\partial_i\LInt(\phi))\|_{H_1} \le
 \br\|\partial_i\LInt(\phi))\|_{L_2} +
 \sum_a \br\|\partial_a(\partial_i\LInt(\phi)))\|_{L_2}
 \le C_R \bigl(\br\|\phi\|_{H_1}^3 + \br\|\phi\|_{H_2}\br\|\phi\|_{H_1}^2\bigr)
\Eeq
with suitable $C_R>0$, which is the estimate \eqref{SlfSmEst} needed.
\end{proof}

\section{Example models}

\subsection{Scalar models}
The simplest examples of classical models are the
scalar models
\Beq
 \cL[\ul\Phi] = -\frac12 \Bigl( \partial_\mu\ul\Phi\partial^\mu\ul\Phi
 + m^2 \ul\Phi^2\Bigr) - \LInt(\ul\Phi)
\Eeq
where $\LInt(\cdot)$ is an entire function on $\mathbb C$ which has an
at least triple zero at the origin
(derivative couplings are physically excluded by Lorentz invariance).

Here is nothing "super", and $M\sol$, if defined, is simply a
real-analytic manifold.

The case of the  $\Phi^4$ model, i.~e.
$\LInt(\ul\Phi) = q\ul\Phi^4$ with $q>0$ and $d=3$,
has already been discussed in \cite{[CAUCHY]}. Using energy conservation,
as in the proof of Cor. \ref{WZSusyCompl} below, the (well-known) completeness
of the $\Phi^4$ model follows from Prop. \ref{SimpSuffCplCrit}(iii).

\subsection{Purely fermionic models}
In the absence of bosonic fields, $N^\Phi=0$, the completeness condition is
void, and Thm. \ref{MainThm} immediately provides a sub-smf of classical
solutions $M\sol\seq M$. In particular, this applies to the Gross-Neveu
models, and to the Thirring model,
which admits a closed formula for $\Xi\sol$ (cf. \cite{[CAUCHY]}).
Note that both $M\sol$ and $M$ are simply points equipped with
infinite-dimensional Grassmann algebras, and the embedding is given by a
surjection $\O(M)\to\O(M\sol)$ ("restriction of field functionals onto
the mass shell").

\subsection{A simple boson-fermion model}\label{Yukawa}
As a more realistic example, we consider the Yukawa
model of meson-nucleon scattering:
The field components are
$\Phi_i$ ($i=1,2,3$) for the three isospin components of the meson field
on the bosonic side and, using complex components,
$\Psi_{\alpha,k},\allowbreak \cj\Psi_{\alpha,k}$ ($k=1,2$, $\alpha=1,\dots,4$)
for the isospin dublet of the nucleon field on the fermionic side.
Thus, the setup is $(d,V)=(4,\mathbb R^{3|16})$.
Using matrix writing for the spinor indices, the Lagrangian is
\Beq
 \cL[\ul\Phi|\ul\Psi] := -\frac\i2
  \br(  \dcj{\ul\Psi}_k\gamma^\mu\partial_\mu\ul\Psi_k -
   \partial_\mu\dcj{\ul\Psi_k}\gamma^\mu\ul\Psi_k
  )
  - \i m^\Psi\dcj{\ul\Psi}_k\ul\Psi_k
  - \frac12 {\partial_\mu\ul\Phi_i\partial^\mu\ul\Phi_i
  - \frac12 (m^\Phi)^2\ul\Phi_i^2}
  - \i g \dcj{\ul\Psi}_k\gamma_5\tau^i_{kl}\ul\Psi_l\ul\Phi_i
\Eeq
where the isospin matrices $\tau^i$ ($i=1,2,3$) are
$\br(\begin{smallmatrix} 0&1\\1&0\end{smallmatrix})$,
$\br(\begin{smallmatrix} 0&-\i\\ \i&0\end{smallmatrix})$, and
$\br(\begin{smallmatrix} 1&0\\0&-1\end{smallmatrix})$, respectively,
$\dcj{\ul\Psi}_{\alpha,k}:=\cj{\ul\Psi}_{\beta,k}\gamma^0_{\beta\alpha}$
is the Dirac conjugate, and $g\in\mathbb R$ is the coupling constant.
The operator $\Dirac$ from \eqref{MassDirOp} is here
$\Dirac=\i(\gamma^\mu\partial_\mu - m^\Psi)$, so that
$K:=-\i(\gamma^\mu\partial_\mu + m^\Psi)$ satisfies \eqref{KDiracKlG}, and
the fermionic influence function $\hA^\Psi$ is now given by \eqref{IAPsi}.

Since the underlying bosonic model describes three free scalar fields,
the model is complete.

\subsection{A supersymmetric model}
Let us consider the most general $d=3$, simply supersymmetric renormalizable
Lagrangian with only chiral superfields, taken from
\cite[(5.12)]{[Wess/Bagger]}. Of course, we use the component field
formulation, with the auxiliary field being eliminated.

The model contains on the bosonic side complex scalar fields
$A_1,\dots,A_n$, and on the fermionic one Weyl spinor fields
$\Psi_1,\dots,\Psi_n$. The parameters are
the "masses" $(m_{ik})\in\mathbb C^{n\times n}$ which are symmetric, the
coupling constants $(g_{ijk})\in\mathbb C^{n\times n\times n}$ which are
symmetric in all indices, and the complex parameters
$(\lambda_k)\in\mathbb C^n$.

For shortness, we set
\Beq
 F_k[\ul A]:= -\lambda_k - m_{ik}\ul A_i - g_{ijk}\ul A_i\ul A_j,\qquad
 \LInt[\ul{\cj A},\ul A] := \cj{F_k[\ul A]} F_k[\ul A].
\Eeq
The Lagrange density is
\Beq
 \cL[\ul{\cj A},\ul A|\;\dcj{\ul\Psi},\ul\Psi] =
  \i \partial_\mu\dcj{\ul\Psi}_i\cj\sigma^\mu\ul\Psi_i
  - \partial_\mu \cj{\ul A}_i\partial^\mu \ul A_i
 - \Re m_{ik}\ul\Psi_i\ul\Psi_k
 - 2 \Re g_{ijk}\ul\Psi_i\ul\Psi_j\ul A_k
 - \LInt[\ul{\cj A},\ul A].
\Eeq
Unfortunately, the $\lambda_k$ spoil the immediate fitting of this Lagrangian
into our model class, since they produce linear terms in the $\ul A_i$.
However, if $c\in\mathbb R^n$ is a critical point of $\LInt$, i. e.
$\partial_{\ul A_i}\LInt[\cj c,c] = \partial_{\ul{\cj A}_i}\LInt[\cj c,c] = 0$,
then we may pass to new variables $\ul A'_i:= \ul A_i - c_i$, getting a new
Lagrangian
\Beq
 \cL'[\ul{\cj A}',\ul A'|\;\dcj{\ul\Psi},\ul\Psi]:=
 \cL[\ul{\cj A}'+\cj c,\ul A'+c|\;\dcj{\ul\Psi},\ul\Psi_i] - \LInt[\cj c,c]
\Eeq
which is easily seen to fit into our model class. (If $\LInt[\cj c,c]\not=0$
then
supersymmetry is spontaneously broken.)

\begin{cor}\label{WZSusyCompl}
This model is complete.
\end{cor}

\begin{proof}
By standard conclusions of QFT, the energy of the underlying bosonic model,
\Beq
 \cH[\ul A']:= \sum\nolimits_i (\partial_0\ul A'_i)^2
  - \cL'[\ul{\cj A}',\ul A'|\;0,0]
 =  \sum\nolimits_i (\partial_0\ul A'_i)^2 +
 \sum\nolimits_{a,i} (\partial_a\ul A'_i)^2  + \LInt[\ul{\cj A'}
  + \cj c,\ul A'+c],
\Eeq
satisfies $\partial_0\cH[A'] =0$ for each solution $A'$ of the underlying
bosonic equations. Since $\LInt[\cj A' + \cj c,A'+c]\ge0$, this implies that
the hypotheses of Prop. \ref{SimpSuffCplCrit}(iii) are satisfied.
\end{proof}

In \cite{[FUNCT]}, we have shown that the usual supersymmetry operators act
as  odd vector fields on the configuration smf $M$.  In the successor paper,
we will show much more: The solution smf $M\sol$ has a natural symplectic
structure, the supersymmetry vector fields restrict to odd vector fields
$Q_\alpha,\cj Q_{\dot\alpha}\in\cX(M\sol)\sodd$ which are just the hamiltonian
vector fields generated by the Noether charges associated to supersymmetry.

\section{Yang-Mills-Dirac-Higgs models in the temporal gauge}
\subsection{The general setting}

In the following, we assume $d\ge1$. Let $G$ be a compact Lie group.
We consider a full Yang-Mills theory coupled to Dirac and Higgs fields, with
the following field components:

-- the gauge potential $A=(A^i_\mu)_{\mu=0,\dots,d}^{i=1,\dots,N^\fg}$ where
$N^\fg$ is the dimension of the Lie algebra $\fg:=\opn lie G$, and the
Lie algebra index $i$ here and in the following always refers to an
orthonormal basis of $\fg$ with respect to a fixed invariant positively
definite scalar product on $\fg$,

-- the Higgs field $\Phi=(\Phi_j)_{j=1,\dots,N^\Phi}$, where $N^\Phi$ is the
dimension of a real orthogonal representation space $V^\Phi$ of $G$,

-- the Dirac field
$\Psi=(\Psi_{k\alpha})_{\alpha=1,\dots,N\D}^{k=1,\dots,N^\Psi}$, where
$N^\Psi$ is the dimension of a unitary representation space $V^\Psi$ of $G$,
and $N\D =2^{1+[d/2]}$ is the dimension of the Dirac representation
$V\D=\mathbb C^{N\D}$ of the Lorentz group.
(We take Dirac spinors for definiteness; everything we do can be carried
over to Weyl, Majorana, and Majorana-Weyl spinors, as far as they are
defined in space dimension $d$.)

Thus, we are given elements $\gamma_\mu\in\opn End (V\D)$ with
$\gamma_\mu\gamma_\nu+\gamma_\nu\gamma_\mu=2g_{\mu\nu}$, which we may assume
to satisfy $\cj\gamma_0\trp = -\gamma_0$, $\cj\gamma_a\trp = \gamma_a$
(of course, this condition is not Lorentz invariant).
As usual, we write the element of $V\D$ as columns, and we define the
{\em Dirac conjugate} of $\psi\in V\D$ as the row $\dcj\psi\in \cj V\D$
given by $\dcj\psi_\alpha:=\cj{\psi_\beta}\gamma^0_{\beta\alpha}$. We get a
Lorentz invariant indefinite pairing
$\cj V\D\times V\D\to\mathbb C$, \ \
$(\dcj\psi,\chi)\mapsto\dcj\psi\chi=
 \cj{\psi_\beta}\gamma^0_{\beta\alpha}\chi_\alpha
$.

We allow the special cases $N^\Phi=0,\ N^\Psi=0$.

We will denote the action of the corresponding
infinitesimal representations of $\fg$ by $\times$, so that
\Beq
 (A_\mu\times \Phi)_k:=  (f^\Phi)_{ik}^l A^i_\mu \Phi_l,\quad
 (A_\mu\times \Psi)_k:= (f^\Psi)_{ik}^l A^i_\mu \Psi_l,
\Eeq
where $(f^\Phi)_{ik}^l,(f^\Psi)_{ik}^l$ are the structure constants of
the corresponding infinitesimal representations.
If $\Phi',\Phi"$ are field monomials with values in $V^\Phi$, and
$\dcj\Psi',\Psi"$ are field monomials with values in $\cj{V^\Psi}\otimes V\D$
and
$V^\Psi\otimes V\D$, respectively, we set
\Beq
 (\Phi'\times\Phi")_i:= (f^\Phi)_{ik}^l \Phi'_k \Phi"_l,\quad
 (\dcj\Psi'\times\Psi")_i:= (f^\Psi)_{ik}^l \dcj{\Psi'_k} \Psi"_l,\quad
\Eeq
where it is understood that orthonormal bases have been used in the
corresponding representation spaces.
Clearly, both field monomials take values in $\fg$.
We will use the usual covariant derivatives
\Beq
 D_\mu\Phi:=\partial_\mu\Phi + A_\mu\times\Phi,\quad
 D_\mu\Psi:=\partial_\mu\Psi + A_\mu\times\Psi.
\Eeq
The standard Lagrangian is
\Beqn YMLagr
  \cL^{\opn YMDH }[\ul A,\ul\Phi|\ul\Psi] := \cL^{\opn YM }[\ul A] +
  \cL^{\opn DH }[\ul A,\ul\Phi|\ul\Psi]
\Eeq
where
\Beq
 \cL^{\opn YM }[\ul A] = -\frac14 F_{\mu\nu}[\ul A]F^{\mu\nu}[\ul A],\qquad
  F_{\mu\nu}[\ul A]
 :=\partial_\mu\ul A_\nu-\partial_\nu\ul A_\mu + \br[\ul A_\mu,\ul A_\nu]
\Eeq
is the  Lagrangian of pure Yang-Mills theory, and
\Beq
 \cL^{\opn DH }[\ul A,\ul\Phi|\ul\Psi] :=
  -\frac\i2 \br(\dcj{\ul\Psi}\gamma^\mu D_\mu\ul\Psi
  - \dcj{D_\mu\ul\Psi}\gamma^\mu\ul\Psi) - \i \dcj{\ul\Psi}m^\Psi\ul\Psi
  -\frac12 D_\mu\ul\Phi_i D^\mu\ul\Phi_i + \LInt(\ul\Phi)
\Eeq
is the Lagrangian of the Higgs field and of minimally coupled Dirac matter.
Here
$m=(m_{ij})_{i,j=1}^{N^\Psi}$ is supposed to be real and $G$-invariant.
(Many textbooks use the opposite Lorentz metric $(1,-1,\dots,-1)$;  since then
$\cj\gamma_0\trp = \gamma_0$, they do not have an imaginary unit in front of
the
fermionic mass term).

For the moment, we suppose for $\LInt$ only that it is a gauge-invariant entire
power series with lower degree $\ge2$.
\Brm
The most popular choice is
\Beq
 \LInt(\ul\Phi)=\frac12 \left(-m^\Phi\ul\Phi_i^2 + h^2(\ul\Phi_i^2)^2\right).
\Eeq
Note that in the underlying free model, the "mass square" $m^\Phi$ is usually
negative,  in order to make $\Phi=0$ not even a local minimum
of the potential. Thus, the "underlying free model" is here
a purely mathematical notion; the free approximation is physically senseful
only at the minima of $\LInt$. Fortunately, we had found before
that scalar fields with negative mass square still behave
classically well.
\Erm

For later use, we recall the Bianchi identity:
\Beqn Bianchi
 D^\mu F^{\nu\lambda} +  D^\nu F^{\lambda\mu} +  D^\lambda F^{\mu\nu} =0.
\Eeq
The field equations resulting from \eqref{YMLagr} are
\begin{subequations}
\Baln  YMEq
  &D^\mu F_{\mu\nu} + J_\nu =0,
\\ \label{EvYMPsi}
  &\br(\gamma^\mu D_\mu + m^\Psi)\Psi =0,
\\  \label{EvPhi}
  &D^\mu D_\mu\Phi_i + \LInt_i(\Phi)  =0,\quad
  \LInt_i(\Phi):=\frac{\partial}{\partial\ul\Phi_i}\LInt(\Phi),
\Eal
\end{subequations}
with the matter current $J_\mu$ given by
\Beq
 J_\mu = \frac{\delta}{\delta\ul A^\mu} \cL^{\opn DH }[A,\Phi|\Psi] =
 -\i\dcj\Psi\times\gamma_\mu\Psi + \Phi \times D_\mu\Phi.
\Eeq
Of course, the corresponding Cauchy problem for the field equations is
not well-posed. There are essentially two main approaches to get a
well-posed Cauchy problem: The simplest possibility is to impose an
explicit gauge condition which diminishes the effective number of degrees
of freedom, like the Lorentz gauge $\partial^\mu A_\mu=0$, or the
{\em temporal gauge} $A_0=0$ (which, however, sacrifices
Poincar\'e invariance).

However, such an explicit gauge breaking is not well adapted to the
needs of quantization. In the physical literature, it is more common
to use a "soft" gauge breaking where the longitudinal degrees of freedom
are not constrained but damped by an additional term in the Lagrangian,
and the correct quantum dynamics is restored by the introduction of
Faddeev-Popov ghost fields. We will follow this approach in section
\ref{SecYMGBrGh}, concentrating in the rest of this section onto the
temporal gauge (despite of its disadvantages, we treat it simply because
for it the completeness result needed is available from
\cite{[Eardley/Moncrief]}, while for the Faddeev-Popov approach the
corresponding result is not yet proven).

In the temporal gauge, the field target space becomes
\Beq
 V:=\rdm\otimes\fg \ \oplus\ V^\Phi\ \oplus\
 \Pi(V^\Psi\otimes_{\mathbb C} V\D),
\Eeq
and we will look for the sub-smf of classical solutions within the smf
of smooth configurations
\Beq
 M\scinf:=\L(\cE\V),\quad  \cE\V:= C^\infty(\rdmm)\otimes V.
\Eeq
(Unfortunately, we cannot treat our favorite quality $\cEc$.
Cf. Rem. \ref{RemConstr}(2) below for the reason why.)

\subsection{Constraint, Cauchy data, and results}

In the temporal gauge $A_0=0$, the
remaining gauge potentials are $(A^i_a)_{a=1,\dots,d}^{i=1,\dots,N^\fg}$.
As usual, we will use the magnetic field strengthes
\Beqn DefBab
 B_{ab}:=F_{ab}=\partial_a A_b - \partial_b A_a -\br[A_a,A_b]
\Eeq
and the electric field strengthes
\Beq
 E_a:=F_{0a}=\partial_0 A_a.
\Eeq
\eqref{YMEq} for $\nu=0$ turns into the constraint
\Beqn DaEaCst
 - D^a E_a + J_0 = 0,\quad\text{i.~e.}\quad
 \partial^a E_a = -\br[A^a,E_a] -\i\dcj\Psi\times\gamma_0\Psi
 + \Phi\times\partial_0\Phi,
\Eeq
while for $\nu=a$ it becomes
\Beqn TGFEs
 \partial_t E_a =\partial^bB_{ba} + \br[A^b,B_{ba}]
  -\i\dcj\Psi\times\gamma_a\Psi + \Phi \times D_a\Phi.
\Eeq

\Brm\relax
\eqref{TGFEs} belongs to the Lagrangian which arises from \eqref{YMLagr}
by setting $\ul A_0:=0$:
\Beqn YMTempG
 \cL^{\opn YMDH-tg }[\ul A] =
 - \frac14 F_{ab}[\ul A]F^{ab}[\ul A] + \frac12 (\partial_0\ul A_a)^2
 + \cL^{\opn DH }[0,\ul A_1,\dots,\ul A_d,\ul\Phi|\ul\Psi].
\Eeq
However, from this point of view, \eqref{DaEaCst} is an "external" constraint.
\Erm

An obvious idea to solve \eqref{DaEaCst} on the Cauchy hyperplane is to
represent  (say) $E_1(0)$ as functional of its restriction $\Erstr$ onto
$x^1=0$ and the remaining Cauchy data. Thus, we take as smf of smooth
Cauchy data
\Baln MParamScinf
 M\param\scinf:=
 &\L( \cE\Cau\otimes\rdm\otimes\fg)_{A\Cau_1,\dots,A\Cau_d}\times
 \L( C^\infty(\mathbb R^{d-1})\otimes\fg)_{\Erstr}\times
 \L(\cE\Cau\otimes\mathbb R^{d-1}\otimes\fg)_{E\Cau_2,\dots,E\Cau_d}\times
  \\
 &\quad\times\L( \cE\Cau\otimes (V^\Phi\oplus V^\Phi))_{\Phi\Cau,\dot\Phi\Cau}
 \times \L( \cE\Cau\otimes \Pi(V^\Psi\otimes_{\mathbb C}
V\D))_{\Psi\Cau},\notag
\Eal
where $\cE\Cau:=C^\infty(\rdm)$.

With the methods of \cite{[CAUCHY]}, \cite{[ABSEVEQ]} one shows:

\begin{lem}
There exists a unique element
\Beq
 E^{\cst}_1=E^{\cst}_1[A\Cau,\Erstr,E\Cau_2,\dots,E\Cau_d,
 \Phi\Cau,\dot\Phi\Cau|\ \Psi\Cau]\in \M^{\cE\Cau\otimes\fg}(M\param\scinf)
\Eeq
such that
\Beq
 \partial^1E^{\cst}_1+[A\Cau_1,E^{\cst}_1] +
 \sum^d_{b=2}\bigl(\partial^b E\Cau_b + \br[A\Cau_b,E\Cau_b]\bigr)
 + \i\dcj\Psi\Cau\times\gamma_0\Psi\Cau - \Phi\Cau\times\dot\Phi\Cau =0,
\Eeq
and $E^{\cst}_1|_{x_1=0} = \Erstr$.
\qed\end{lem}

\Brmn RemConstr
(1) It easily follows that the element
\Beqn CauCst
 \partial^a E\Cau_a +[A\Cau_a,E\Cau_a]  + \i\dcj\Psi\Cau\times\gamma_0\Psi\Cau
 -\Phi\Cau\times\dot\Phi\Cau \in \M^{\cE\Cau\otimes\fg}(M\Cau\scinf)
\Eeq
cuts out a sub-supermanifold (cf. \cite[2.12]{[WHAT]})
$M^{\opn Cau,cst }\scinf \seq M\Cau\scinf$ in the smf of unconstrained smooth
Cauchy data
\Bal
 M\Cau\scinf :=
 &\L( \cE\Cau\otimes\rdm\otimes\fg)_{A\Cau}\times
  \L( \cE\Cau\otimes\rdm\otimes\fg)_{E\Cau}\times
 \\
 &\quad\times
 \L( \cE\Cau\otimes (V^\Phi\oplus V^\Phi))_{\Phi\Cau,\dot\Phi\Cau}\times
 \L( \cE\Cau\otimes \Pi(V^\Psi\otimes_{\mathbb C} V\D))_{\Psi\Cau}.
\Eal
Indeed, the sub-smf sought is just the image of the morphism
$M\param\scinf\to M\Cau\scinf$,
the pullback of which maps $E\Cau_1$ to $E^{\cst}_1$, and all other
coordinate components  to the coordinate components with the same name.

(2) Unfortunately, the analogous assertion for our favorite quality
$\cEc$ seems to be false; at any rate, we will have in general
$E^{\cst}_1\not \in \M^{\cEc\Cau\otimes\fg}(M\param)$ where $M\param$ is
formed like \eqref{MParamScinf} but using $\cEc\Cau:=\cD(\rdm)$ instead of
$\cE\Cau$, and $C^\infty_0(\mathbb R^{d-1})$ instead of
$C^\infty(\mathbb R^{d-1})$.
\Erm

The main result of this section is:

\begin{thm}\label{YMTGThm}
\begin{itemize}
\item[(i)]  Let be given a point $p=(s,x)\in\rdmm$ with $s\not=0$ and
two $Z$-families $(A',\Phi'|\Psi'),\ \ (A",\Phi"|\Psi")$  of
\eqref{YMEq}--\eqref{EvPhi} which satisfy $A'_0=A"_0=0$.  Suppose that,
with the notations of \ref{CauUnCaus},
\Bal
  &\br(D_{A'}^\mu F_{\mu\nu}[A'] + J_\nu[A',\Phi'|\Psi'])|_{\Omega(p)} =0,
\quad
  &&\br({\gamma^\mu (D_{A'})_\mu + m^\Psi}) \Psi'|_{\Omega(p)} = 0,
 \\
  &\br({D_{A'}^\mu (D_{A'})_\mu\Phi'_i + \LInt_i(\Phi')})|_{\Omega(p)}= 0,
\Eal
and analogously for $(A",\Phi"|\Psi")$, and that, writing
$\Delta=(A'-A",\Phi'-\Phi")$, we have
\Beq
 \Delta(0)|_{\J(p)}=0,\quad
 \partial_0\Delta(0)|_{\J(p)}=0,\quad
 (\Psi'-\Psi")(0)|_{\J(p)}=0.
\Eeq
Then
\Beq
  \br(A'-A",\Phi'-\Phi"|\Psi'-\Psi")|_{\Omega(p)}=0.
\Eeq

\item[(ii)] Suppose that $d\ge2$, and that there exists a Sobolev index $k>d/2$
such that for every solution
\Beq
 (a,\phi) \in C^\infty((t_0,t_1)\times\rdm)\otimes(\rdmm\otimes\fg \oplus
V^\Phi)
\Eeq
of the Yang-Mills-Higgs equations in the temporal gauge,
\Beq
  D^\mu F_{\mu\nu}[a] + \phi\times D_\nu \phi =0,\quad
  D^\mu D_\mu \phi_i + \LInt_i(\phi)  =0, \quad a_0=0,
\Eeq
on an open time interval $(t_0,t_1)$  such that
$\supp(a,\phi)(t)$ is compact for some (and hence all) $t\in(t_0,t_1)$, we have
\Beq
 \sup_{t\in(t_0,t_1)} \max
 \br\{
 \| a^i_\mu(t)\|_{H_{k+1}(\rdm)},\
 \|\phi_j(t)\|_{H_{k+1}(\rdm)},\
 \|\partial_t a^i_\mu(t)\|_{H_{k}(\rdm)},\
 \|\partial_t \phi_j(t)\|_{H_{k}(\rdm)}
 \} < \infty
\Eeq
where $i=1,\dots,N^{\fg}$, $\mu=1,\dots,d$, $j=1, \dots,N^\Phi$. Then the
following assertions are valid.

\begin{itemize}
\item[(1)] There exists a unique solution family
\Beqn YMTGUniv
\begin{split}
 \Xi\sol&=(A\sol,\Phi\sol|\Psi\sol)
\\
 &=\Xi\sol[A\Cau,\Erstr,E\Cau_2,\dots,E\Cau_d,\Phi\Cau,\dot\Phi\Cau|\Psi\Cau]
 \in \M^{\cE\V}(M\param\scinf)
\end{split}
\Eeq
of \eqref{YMEq}--\eqref{EvPhi} which satisfies $A\sol_0=0$ with the
initial data
\Bal
 A_a\sol(0,\cdot) &=A\Cau_a(\cdot),    \quad&& a=1,\dots,d,
 \\
 \partial_t A_1\sol(0,0,x_2,\dots,x_d) &= \Erstr(x_2,\dots,x_d),
 \\
 \partial_t A_a\sol(0,\cdot) &= E\Cau_a(\cdot), \quad&& a=2,\dots,d,
 \\
 (\Phi\sol,\partial_t\Phi\sol|\Psi\sol)(0,\cdot)
 &=(\Phi\Cau,\dot\Phi\Cau|\Psi\Cau)(\cdot).
\Eal

\item[(2)] This solution family is universal, i.~e. any other smooth
solution family $(A',\Phi'|\Psi')\in \M^{\cE\V}(Z)$ of
\eqref{YMEq}--\eqref{EvPhi} which satisfies $A'_0=0$ is in a unique way
a pullback of \eqref{YMTGUniv}.

\item[(3)] The image of $M\param\scinf\too{\Xi\sol} M\scinf$ is a sub-smf.
\end{itemize}

\item[(iii)] In particular, suppose that either

\begin{itemize}
\item[(4)]
$d=2$, and $\LInt = V(\ul\Phi_i\ul\Phi_i)$ with an entire power series
$V(r)$ of polynomial growth which satisfies $V(0)=0$ and $V(r)\ge - Cr$
for suitable $C>0$, or
\item[(4)]
 $d=3$, and $\LInt$ is a $G$-invariant polynomial in $\ul\Phi$
of degree $\le4$.
\end{itemize}
Then the assertions (1), (2), (3) are true.
\end{itemize}
\end{thm}

We postpone the proof of the Theorem to the next section.

\subsection{The first-order system}\label{TG1stOrder}

Let us shortly discuss the obstackles to a naive approach.
The quadratic part of \eqref{YMTempG} has not the required form
\eqref{GenFrmOfLKin}. A straightforward  generalization of our previous
approach to \eqref{YMTempG} fails for the following reason:
The linearized equation of motion for $A$ is
\Beq
 K^{ai,bj} A_{aj} = 0,\quad
 K^{ai,bj} = \delta_{ij}(-\delta^{ab} \KlGord + \partial^a\partial^b).
\Eeq
The corresponding Fourier-transformed influence function $\hat\cA(t,p)$
is given as $\hat\cA(t,p)= 1_{m\times m} \otimes \hat\cA'(t,p)$ where
$\hat\cA'(t,p)$ is determined by
\Beq
 0 = \partial_t^2 \hat\cA' + (p^2 1_{d\times d} - pp\trp)\hat\cA',\quad
 \hat\cA'(0)=0,\ \ \partial_t\hat\cA'(0)=(2\pi)^{-d/2}1_{d\times d}.
\Eeq
Explicitly,
\Beq
 (2\pi)^{d/2}\hat\cA'(t,p)^a_c =
 \frac{\sin \br|p|t}{\br|p|} \br(\delta^a_c - \frac{p^ap_c}{\br|p|^2})
 + t \frac{p^ap_c}{\br|p|^2}
\Eeq
Thus, \eqref{TheCAEst0} (with $\hat\cA^\Phi:=\hat\cA$)
is violated, since the spatially longitudinal  direction is not sufficiently
smoothened, and we cannot assign to $A$ the smoothness offset $1$,
which we need because of the derivative coupling implicitly arising in
\eqref{YMEq}. (In another language, this obstackle to a naive approach was
observed earlier, cf. e.g. \cite{[Eardley/Moncrief]}.)

In order to get a first-order system which fits into the class considered
in \cite{[ABSEVEQ]}, we follow Segal's idea and temporarily forget the
definition of the magnetic field strengthes \eqref{DefBab}, considering them
as independent $\fg$-valued antisymmetric tensor field $B_{ab}$ instead.
Also, we introduce a new $\fg$-valued scalar field $L$ for the longitudinal
component
\Beqn CstL
 L=\partial^aA_a
\Eeq
(otherwise, the term $(\partial^aA_a)\Phi$ would lead to smoothness
difficulties). Thus, we consider the following system in the commuting
fields $A_a,E_a,B_{ab},L,\Phi,\dot\Phi$ and the anticommuting fields
$\Re\Psi,\Im\Psi$:
\begin{subequations}
\Baln  ExtSyst
 \partial_t A_a &= E_a, &&
 \\
 \label{tgDtEa}
 \partial_t E_a &=\partial_bB_{ba} + \br[A_b,B_{ba}] + J_a, &&
 \\
 \label{tgTDConstr}
 \partial_t B_{ab}
  &=\partial_a E_b - \partial_b E_a + \br[A_a,E_b] - \br[A_b,E_a]
 && \text{(by \eqref{Bianchi}),}
 \\
 \label{tgEvL}
 \partial_t L&= -[A_a,E_a] + J_0
 && \text{(by \eqref{DaEaCst}),}
 \\ \label{TGPtPhi}
 \partial_t\Phi &=\dot\Phi, &&
 \\ \label{TGPtdPhi}
 \partial_t\dot\Phi&=\Delta\Phi + L\times\Phi + 2A_a\times\partial_a\Phi
 + A_a\times(A_a\times\Phi) + \LInt_*(\Phi)
 && \text{(by \eqref{EvPhi}, \eqref{CstL}),}
 \\\label{ExtSysLast}
 \partial_t\Psi &= - (\gamma^0)^{-1} \br(\gamma^a D_a +  m^\Psi)\Psi
 && \text{(by \eqref{EvYMPsi}),}
\Eal
\end{subequations}
where we have abbreviated
\Beq
 \LInt_*(\Phi):=(\LInt_i(\Phi))_{i=1}^{N^\Phi},\quad
 J_0 := - \i\dcj\Psi\times\gamma_0\Psi + \Phi\times\dot\Phi,\quad
 J_a := -\i\dcj\Psi\times\gamma_a\Psi + \Phi\times D_a\Phi.
\Eeq
We claim that this system together with the constraints \eqref{DefBab},
\eqref{DaEaCst}, \eqref{CstL} is equivalent with the original system
\eqref{YMEq}--\eqref{EvPhi} together with the constraint $A_0=0$:

\begin{lem}\label{LemTmpGaugeEv}
(i)
Let be given a solution family $(A,\Phi|\Psi)$ of \eqref{YMEq}--\eqref{EvPhi}
which satisfies $A_0=0$.
Then
\Beqn ExtTGConfFam
 (A_a,E_a, B_{ab},L,\Phi,\dot\Phi|\ \Re\Psi,\Im\Psi)
\Eeq
where $E_a:=\partial_tA_a$,\ \  $B_{ab}:=F_{ab}[0,A_1,\dots,A_d]$,\ \
$L:=\partial^aA_a$, and $\dot\Phi:=\partial_0\Phi$,
is a solution family of the system \eqref{ExtSyst}--\eqref{ExtSysLast}
which satisfies the constraints
\eqref{DefBab}, \eqref{DaEaCst}, and \eqref{CstL}.

(ii) Conversely, let be given a solution family \eqref{ExtTGConfFam} of
\eqref{ExtSyst}--\eqref{ExtSysLast} which satisfies the constraints
\eqref{DefBab}, \eqref{DaEaCst}, and \eqref{CstL} at $t=0$.
Then these constraints will be satisfied at all times, and
$(A,\Phi|\Psi)$ will be a solution family of \eqref{YMEq}--\eqref{EvPhi}.

(iii) The system  \eqref{ExtSyst}--\eqref{ExtSysLast}  belongs to the class
considered in \cite{[CAUCHY]}, \cite{[ABSEVEQ]}, with assigning the
smoothness offset $1$ to $\Phi_i$, and $0$ to all remaining field components.
Moreover, this system  is causal in the sense of \cite{[CAUCHY]},
\cite{[ABSEVEQ]}.
\end{lem}

\begin{proof}
Ad (i). This follows from the construction.

Ad (ii). Using \eqref{tgTDConstr} and \eqref{Bianchi}, we have
\Beq
 \partial_t\br(B_{ab} - F_{ab}[0,A_a]) =0,
\Eeq
which implies that \eqref{DefBab} is satisfied for all times. Using
\eqref{tgDtEa} we find
\Beq
\begin{split}
  &\partial_t\br(D^a E_a - J_0 ) =
  D^a(D^bB_{ba} +J_a)  -\partial_tJ_0 = D^\mu J_\mu
\\
  & = \i\dcj{D^\mu \gamma_\mu \Psi}\times\Psi
    - \i\dcj\Psi\times D^\mu \gamma_\mu\Psi
    + D^\mu\Phi \times  D_\mu\Phi + \Phi \times D^\mu D_\mu\Phi.
\end{split}
\Eeq
Now \eqref{TGPtdPhi} yields
$D^\mu D_\mu\Phi + (L-\partial^aA_a)\times\Phi +  \LInt_*(\Phi)  =0$;
on the other hand, infinitesimal gauge invariance of $\LInt$ is equivalent
with $\ul\Phi\times \LInt_*(\ul\Phi)=0$.
Since $G$ acts orthogonally on $V^\Phi$  we have
$(f^\Phi)_{ik}^l=-(f^\Phi)_{il}^k$ and hence
$D^\mu\Phi\times D^\mu \Phi=0$. Investing all this and
also \eqref{ExtSysLast}, we get
\Beqn DtCst
\begin{split}
  &\partial_t\br(D^a E_a - J_0 )
  = -\i\dcj{m^\Psi\Psi}\times\Psi + \i\dcj\Psi\times m^\Psi\Psi
 + \Phi\times (\br({(\partial^aA_a-L)\times\Phi}) -  \LInt_*(\Phi) )
\\
  &= \Phi\times \br({(\partial^aA_a-L)\times\Phi}).
\end{split}
\Eeq
Using \eqref{tgEvL}, \eqref{ExtSyst} we have
\Beqn PtLAa
  \partial_t\br(L-\partial^aA_a) = J_0 - D^a E_a,
\Eeq
and hence
\Beqn Pt2LAa
  \partial_t^2\br(L-\partial^aA_a) =
  \Phi\times \br({(L-\partial^aA_a)\times\Phi}).
\Eeq
By hypothesis, $\br(L-\partial^aA_a)(0)=0$ and $\br(J_0 - D^a E_a)(0)=0$,
which by \eqref{PtLAa} yields $\partial_t\br(L-\partial^aA_a)(0)\allowbreak=0$.
Therefore, \eqref{Pt2LAa} implies that \eqref{CstL} is satisfied for all times.
Now \eqref{DtCst} together with the initial data yields that the constraint
\eqref{DaEaCst} is satisfied at all times, too. The remaining field equations
are now obviously satisfied.

Ad (iii). For finding the influence function $\hA$ of the free evolution,
we can, of course, consider the sectors separately:
$\hA=\opn diag (\hA^{AEB},\hA^L,\hA^\Phi,\hA^\Psi)$.
Obviously, $\hA^L=(2\pi)^{-d/2}$ is constant (here and below, we do not
indicate the Kronecker deltas for the colour indices). $\hA^\Phi$ is given
by \eqref{IAPhi} with
$m^\Phi_{ij}:=
\frac{\partial^2}{\partial\ul\Phi_i\partial\ul\Phi_j}\LInt(0)/2$,
and $\hA^\Psi=A^\Psi$ is the same as in \ref{Yukawa}. Finally, the entries of
\Beq
\hA^{AEB} =\begin{pmatrix}
(2\pi)^{-d/2}\delta^a_c &  \hX^a_c &  \hX^{ab}_c \\
0& \hA^a_c &  \hA^{ab}_c\\
0& \hA^a_{cd}&   \hA^{ab}_{cd}
\end{pmatrix}
\Eeq
are determined by the spatially Fourier-transformed free field equations
\Beq
 \partial_t \hX^I_c = \hA^I_c,\quad
 \partial_t \hA^I_c = \i p_d\hA^I_{dc},\quad
  \partial_t \hA^I_{cd}   =\i p_c\hA^I_d - \i p_d\hA^I_c,
\Eeq
with $I$ standing for $a$ or $ab$, and the initial values
$\hA^{AEB}(0) =(2\pi)^{-d/2}\opn diag
(\delta^a_c,\delta^a_c ,\delta\Bb a{[c}\delta\Bb b{d]})$
(here and in the following, we use Bach's antisymmetrizing brackets).
By direct verification, we have
\Beqn hAAEB
\hA^{AEB} =(2\pi)^{-d/2}\begin{pmatrix}
 \delta^a_c
&
 {\displaystyle \frac{\sin \br|p|t}{\br|p|}
 \br(\delta^a_c - \frac{p^ap_c}{\br|p|^2}) + t \frac{p^ap_c}{\br|p|^2}}
&
 {\displaystyle \i \frac{1-\cos\br|p|t}{\br|p|^2}p^{[a}\delta^{b]}_c}
\\
 0
&
 {\displaystyle \cos\br|p|t\br(\delta^a_c - \frac{p^ap_c}{\br|p|^2})
  + \frac{p^ap_c}{\br|p|^2}}
&
 {\displaystyle \i\frac{\sin \br|p|t}{\br|p|} p\Bb{[a}{} \delta\Bb{b]}c }
\\
0
&
 {\displaystyle \i\frac{\sin \br|p|t}{\br|p|} p\Bb{}{[c}\delta\Bb a{d]}}
&
 {\displaystyle \delta^a_{[c}\delta^b_{d]} + \frac{\cos\br|p|t-1}{\br|p|^2}
 \br(p^ap\Bb {}{[c}\delta \Bb b{d]} - p^bp\Bb{}{[c}\delta\Bb a{d]})
 }
\end{pmatrix}
\Eeq

(with the continuous extension at $p=0$ understood). Since each entry remains
bounded for $\br|p|\to\infty$, the smoothness conditions are satisfied.

The causality assertion follows from the Paley-Wiener Theorem.
\end{proof}

\begin{proof}[Proof of Thm. \ref{YMTGThm}]
Ad (i). This follows from the Lemma and the results of \cite{[ABSEVEQ]}.

Ad (ii).  Let
\Beq
 V\ext := \fg\otimes\br(\rdm \oplus \rdm \oplus \Lambda^2\rdm \oplus \mathbb R)
 \oplus V^\Phi \oplus V^\Phi \ \oplus \ \Pi(V^\Psi\otimes_{\mathbb C} V\D)
\Eeq
be the field target space for the extended system
\eqref{ExtSyst}--\eqref{ExtSysLast}, so that the corresponding
smf of smooth Cauchy data is $M^{\opn extCau }\scinf=\L(\cE\Cau\otimes V\ext)$.
Consider the smf morphism
\Beqn MPar2MExtCau
 M\param\scinf
 \too{(A_a\Cau,\Erstr,E\Cau_2,\dots,E\Cau_d,
  B_{ab}\Cau:=F_{ab}[0,A_1\Cau,\dots,A_d\Cau],
  L:=\partial^a A\Cau_a,
  \Phi\Cau, \dot\Phi\Cau|\Psi\Cau)}
 M^{\opn extCau }\scinf.
\Eeq
Let $Y$ denote temporarily the image of its underlying map, i.~e. the
set of all bosonic Cauchy data
\Beq
 (a_c,e_c, b_{cd},l,\phi,\dot\phi)\in\cE^{\opn V \ext}\seven
\Eeq
which satisfy the constraints
\begin{align}
 b_{cd}&=\partial_c a_d - \partial_d a_c
 -\br[a_c,a_d],\quad  l=\partial^c a_c,
 \notag
 \\ \label{Ecst}
 \partial^c e_c &= -[a_c,e_c] + \phi\times\dot\phi.
\end{align}
Now the hypothesis just says that the system
\eqref{ExtSyst}--\eqref{ExtSysLast}
is $(\cH^{\opn V \ext_k},Y \cap (\cE_c^{\opn V \ext})\seven )$-complete,
where we use the Sobolev space $\cH^{\opn V \ext_k}$ as defined in
\cite{[CAUCHY]}, \cite{[ABSEVEQ]}.

On the other hand, in order to apply \cite[Thm. 3.4.5]{[CAUCHY]}, we have
to show that each bosonic Cauchy datum from $Y$ is approximable in the sense
of \cite[3.4]{[CAUCHY]}, i.e. for each ball $n\mathbb B$
(where $\mathbb B:=\{x\in \rdm:\ \ \br|x|\le 1\}$)
there exist bosonic Cauchy data
$(a'_c,e'_c, b'_{cd},l',\phi',\dot\phi')\in Y \cap (\cE_c^{\opn V \ext})\seven$
which coincide on $n\mathbb B$ with the Cauchy data given, and which have
infinite lifetime. Indeed, this follows easily from the hypothesis and the
following Lemma:

\begin{lem}\label{RstInitDta}
Suppose $d\ge2$, and let be given $C^\infty$ bosonic initial data
$(a,e,\phi,\dot\phi)\in \cE\V\seven$ which satisfy the constraint \eqref{Ecst}.
For given $n>0$, there exist compactly supported bosonic initial data
\Beqn LocIDta
 (a',e',\phi',\dot\phi')\in (\cEc\V)\seven
\Eeq
which satisfy
\Baln SameOnBall
 &(a',e',\phi',\dot\phi')|_{n\mathbb B} = (a,e,\phi,\dot\phi)|_{n\mathbb B},
 \\\label{cst'}
 &\partial^c e'_c = -[a'_c,e'_c] + \phi'\times\dot\phi'.
\Eal
\end{lem}

\begin{proof}
Choose a buffer function $g\in C^\infty(\rdm)$ such that $g|_{n\mathbb B} =1$
and $g|_{\rdm \setminus (n+1)\mathbb B} =0$.
The idea is to get rid of the charge
$(-1)^c \int_{n\mathbb S} dx_1\dots dx_{c-1}dx_{c+1}\dots dx_d ge_c$
by adding the opposite charge on a  of the ball $n\mathbb B$.
Let $p:=(0,3n,0,\dots,0)\in\rdm$, and let
\Bal
 &a'(x):=(ga)(x) + (ga)(x-p),
 \\
 &e"_2(x):=(ge_2)(x) - (ge_2)(x-p),\quad
 &&e'_c(x):=(ge_c)(x) - (ge_c)(x-p) \quad \text{for $c=3,\dots,d$,}
 \\
 &\phi'(x):=(g\phi)(x) - (g\phi)(x-p)),\quad
 &&\dot\phi'(x):=(g\dot\phi)(x) + (g\dot\phi)(x-p).
\Eal
In order to determine $e'_1,e'_2$, choose another buffer function
$h\in C^\infty(\mathbb R)$ with
$h|_{[-n-1,n+1]} = 1$, \ \ $h|_{\mathbb R\setminus[-n-2,n+2]} = 0$.
Let $f\in C^\infty(\rdm)\otimes V^\fg$ be the solution of the ODE
\Beqn SetF
  \partial_1 f = -[a'_1,f] - \partial_2 e"_2 - [a'_2,e"_2]
 - \sum_{c=3}^d \br(\partial_c e'_c + [a'_c,e'_c])
 + \phi'\times\dot\phi'
\Eeq
with the initial data ($x=(x_1,x_2,\ul x)$)
\Beq
 f(0,x_2,\ul x):= (ge_1)(0,x_2,\ul x) - (ge_1)(0,x_2-3n,\ul x)
\Eeq
and let $e'_1(x):=h(x_1)f(x)$. Let
\Beqn SetE2
 e'_2(x):= e"_2(x) - \partial_1h(x_1) \int_{-\infty}^{x_2}
 d\xi_2 f(x_1,\xi_2,\ul x)
\Eeq
Except for the component $e'_1$, the validity of \eqref{SameOnBall} is clear
by construction, while $(e'_1-e_1)|_{n\mathbb B}=0$ follows since
$e'_1,e_1$ satisfy on $n\mathbb B$ the same ordinary differential equation with
the same initial data; this proves \eqref{SameOnBall}.

With a similar argument, we have $f(x)=-f(x-p)$ for $x_2\in[2n-1,4n+1]$.
Note also that $f(x)=0$ if
$x\in\mathbb R \setminus \br([-n-1,n+1]\cup[2n-1,4n+1])$,
so the integral in \eqref{SetE2} is well-defined, and becomes zero for
$x_2\in \mathbb R \setminus[-n-1,4n+1]$. This guarantees
that $e'_2$ has compact support, and \eqref{LocIDta} is now easily verified.

It remains to show \eqref{cst'}. For $x_1\in[-n-1,n+1]$ we have $e'_1=f$ and
$e'_2= e"_2$, so that \eqref{SetF} yields \eqref{cst'}. For
$x_1\in\mathbb R\setminus[-n-1,n+1]$, \eqref{cst'} simplifies to
$\partial_1 e'_1 + \partial_2 e'_2=0$, which is fulfilled by construction.
\end{proof}

By \cite[Thm. 3.4.5]{[CAUCHY]}, there now exists a solution family
\Beq
 \Xi\extsol
 =(A_a\extsol,E_a\extsol,B_{ab}\extsol,L\extsol,
  \Phi\extsol,\dot\Phi\extsol|\Psi\extsol)
\Eeq
parametrized by $M\param\scinf$ which has \eqref{MPar2MExtCau} as its Cauchy
data.
Now  $(A_a\extsol,\Phi\extsol|\Psi\extsol)\in \M^{\cE\V}(M\param\scinf)$
will be the universal solution family wanted.
For showing that  its image will be a sub-smf, the proof of
\cite[Thm. 2.4.1]{[ABSEVEQ]} carries over.

Ad (iii). This follows from the results of Ginebre/Velo \cite{[Ginebre/Velo]}
and Eardley/Moncrief \cite{[Eardley/Moncrief]}.
\end{proof}

\Brm
(1) Cf. also \cite{[ChoYM]}, where, however, the global existence is proved
only for sufficiently small Cauchy data and coupling constants.

(2) In $d=1$, Lemma \ref{RstInitDta} may become wrong (take e.g. $G=U(1)$ with
$V^\Phi=0$; then the constraint is simply $\partial_1e_1=0$). The best we can
say is:

Suppose that there exists some $k>1/2$ such that for every $n>0$ and every
choice of smooth bosonic Cauchy data
$\xi=(a_1,e_1,\phi,\dot\phi)$ which satisfy the constraint
$D^1e_1=\phi\times\dot\phi$, there exists a bosonic solution
$(a'_1,\phi)\in
 C^0((-n,n),\ H_{k+1}(\mathbb R)\otimes(\fg \oplus V^\Phi))\cap
 C^1((-n,n),\ H_k(\mathbb R)\otimes(\fg \oplus V^\Phi))
$
of
\Beq
  D^1 e_1 = \phi' \times \partial_0\phi',\quad
  \partial^0 e_1 = \phi' \times D_1\phi',\quad
  D^\mu D_\mu \phi'_i + \LInt_i(\phi')  =0,
\Eeq
the Cauchy data of which agree with $\xi$ on the ball $n\mathbb B$.
Then the assertions of Thm. \ref{YMTGThm}.(ii).(1-3) are valid.
\Erm

\section{Yang-Mills-Dirac-Higgs models in the Faddeev-Popov approach}
\label{SecYMGBrGh}

\subsection{The Lagrangian}
The approach of explicitly stating the gauge condition like $A_0=0$
is known to have some defects, like the uncertainty whether the results
will be still Poincar\'e invariant. In fact, in almost all calculations of
experimentally verifiable predictions, one introduces a gauge symmetry
breaking term, and Faddeev-Popov ghosts. That is, one introduces new
fermionic fields $\Ups_i,\Lambda_i$ ({\em Faddeev-Popov ghost and antighost
fields}) with $i=1,\dots,N^{\fg}$ which transform under the adjoint
representation of the gauge group and are scalars with respect to the
Poincar\'e group (so they are the "ghost drivers" with respect to the
Pauli Theorem on the connection between spin and statistics). Thus
\Beq
 D^\mu\Ups= \partial^\mu\Ups + [A^\mu,\Ups],\quad
 D^\mu\Lambda= \partial^\mu\Lambda + [A^\mu,\Lambda].
\Eeq
One now works with the Lagrangian
\Beqn YMGB
  \cL^{\opn YMDH-gb }[\ul A,\ul\Phi|\ul\Lambda,\ul\Ups,\ul\Psi]
   := \cL^{\opn YMDH }[\ul A,\ul\Phi|\ul\Psi]
   - \frac\zeta2 \Bigl( \partial^\mu\ul A_\mu\Bigr)^2
  + \i\partial_\mu\ul\Lambda_i D^\mu\ul \Ups_i
\Eeq
(cf. \eqref{YMLagr}) where the parameter $\zeta\in\mathbb R\setminus0$
is arbitrary.

The field equations resulting from \eqref{YMGB} are
\begin{subequations}
\Baln YMGBEq
  &D^\mu F_{\mu\nu} + \zeta \partial_\nu (\partial^\mu A_\mu) + J_\nu =0,
 \\ \label{EvYMUps}
 &\partial^\mu D_\mu\Ups =0,
 \\ \label{EvYMAgh}
 &D_\mu\partial^\mu \Lambda =0,
 \\ \label{YMFPPsi}
  &\br(\gamma^\mu D_\mu + m^\Psi)\Psi =0,
\\  \label{YMGBEqLast}
  &D^\mu D_\mu\Phi + \LInt_*(\Phi)  =0,
\Eal
\end{subequations}
with the matter current now given by
\Beqn FPMCur
 J_\mu = \frac{\delta}{\delta\ul A^\mu}
 \br( \partial_\nu\ul\Lambda_i D^\nu\ul \Ups_i + \cL^{\opn DH })[A,\Phi|\Psi]
  = - \i\br[\partial_\mu\Lambda,\Ups] + \Phi \times D_\mu\Phi
   - \i\dcj\Psi\times\gamma_\mu\Psi.
\Eeq
In this approach, all the difficulties connected with constraints
disappear: The smfs of Cauchy data and the smf of the configuration
are given by
\Bal
 M\Cau:=
 &\L( \cEc\Cau\otimes\rdmm\otimes\fg)_{A\Cau_0,\dots,A\Cau_d}\times
 \L(\cEc\Cau\otimes\rdmm\otimes\fg)_{\dot A\Cau_0,\dots,\dot A\Cau_d}\times
 \\
 &\quad\times\L( \cEc\Cau\otimes (V^\Phi\oplus V^\Phi))_{\Phi\Cau,\dot\Phi\Cau}
 \times
 \L(\cEc\Cau\otimes \Pi(\fg\oplus\fg\oplus\fg\oplus\fg))
   _{\Lambda\Cau,\dot\Lambda\Cau,\Ups\Cau,\dot\Ups\Cau}
 \times
 \\
 &\quad\times\L(\cEc\Cau\otimes \Pi(V^\Psi\otimes_{\mathbb C} V\D))_{\Psi\Cau},
 \\
 M:=
 &\L(\cEc\V) = \L( \cEc\otimes\rdmm\otimes\fg)_{A,\dots,A_d}\times
 \L( \cEc\otimes V^\Phi)_{\Phi\Cau}\times
 \\
 &\quad\quad\quad\times\L(\cEc\otimes \Pi(\fg\oplus\fg))_{\Lambda,\Ups}
 \times
 \L( \cEc\otimes \Pi(V^\Psi\otimes_{\mathbb C} V\D))_{\Psi\Cau}.
\Eal

\Brm
(1) The resulting description of the configuration space of such a model
as supermanifold was first given in \cite{[Abramov 86]} (although in
another technical setting).

(2) Let us comment about the imaginary unit in front of the ghost term,
which is absent in most textbook presentations, but which in our approach
is necessary in order to ensure the reality of $\cL^{\opn YMDH-gb }$. In
fact, there is some confusion in literature about whether $\Lambda$ and
$\Ups$ should be mutually conjugated complex fields or independent
real (or, in textbook language, hermitian) fields. One can work with
either of both interpretations since if $\Delta(A)$ denotes the Faddeev-Popov
determinant the formal path integral representation
\Beq
 \Delta(A) = \int [d\Lambda][d\Ups]
 \exp\Bigl(\i \int d^{d+1}x\partial_\mu\Lambda_i(x) D^\mu\Ups_i(x) \Bigr)
\Eeq
is valid for both interpretations. However, the next step in the textbooks is
to include the term $\partial_\mu\ul\Lambda_i D^\mu\ul \Ups_i$ into the
Lagrangian, without bothering about its reality.

Here we settle for treating $\Lambda$ and $\Ups$ as independent real
fields. The additional $\i$ introduces an inessential constant factor
$\i^\infty$ into the functional integral. Also, the perturbation theory
remains essentially unchanged: both the $\partial_\mu\Lambda_i A^\mu\Ups_i$
vertex and the kinetic operators for $\Lambda,\Ups$ get an additional factor
$\i$, and hence the Green functions of $\Lambda,\Ups$ get a factor $1/\i$.
Since ghosts appear only in closed loops, these additional factors cancel away.
(Cf. also \cite{[AchPelm]} as one of the very few exceptions in the textbook
literature; this book considers also the canonical quantization of the ghosts
and therefore uses the factor $\i$, too.)
\Erm

\subsection{The main result}

\begin{thm}\label{YMFPThm}
\begin{itemize}
\item[(i)]
Let be given a point $p=(s,x)\in\rdmm$ with $s\not=0$ and two $Z$-families
$(A',\Phi'|\ \Lambda',\Ups',\Psi'),\allowbreak (A",\Phi"|\
\Lambda",\Ups",\Psi")$.
Suppose that
\Bal
  &\left(D_{A'}^\mu F_{\mu\nu}[A'] + \zeta \partial_\nu (\partial^\mu A'_\mu)
  + J_\nu[A',\Phi'|\Psi']\right)|_{\Omega(p)} =0,
  &&\bigl(\partial_\mu D_{A'}^\mu\Ups'\bigr)|_{\Omega(p)}=
  \bigl(D_{A'}^\mu\partial_\mu \Lambda'\bigr)|_{\Omega(p)} =0,
  \\
  &\br({D_{A'}^\mu (D_{A'})_\mu\Phi'_i + \LInt_i(\Phi')})|_{\Omega(p)}= 0,
  &&\br({\gamma^\mu (D_{A'})_\mu + m^\Psi})\Psi'|_{\Omega(p)} = 0,
\Eal
and analogously for $(A",\Phi"|\ \Lambda",\Ups",\Psi")$, and that, setting
$\Delta=(A'-A",\Phi'-\Phi"|\ \Lambda'-\Lambda",\Ups'-\Ups")$, we have
\Beq
 \Delta(0)|_{\J(p)}=0,\quad \partial_0\Delta(0)|_{\J(p)}=0,
 \quad (\Psi'-\Psi")(0)|_{\J(p)}=0.
\Eeq
Then
$\br(A'-A",\Phi'-\Phi"|\ \Lambda'-\Lambda",\Ups'-\Ups",\Psi'-\Psi")
 |_{\Omega(p)}=0
$.

\item[(ii)] Suppose there exists a Sobolev index $k>d/2$ such that
for every bosonic solution
$(a,\phi) \in C^\infty((t_0,t_1)\times\rdm)\otimes(\fg\otimes\rdmm
\ \oplus\ V^\Phi)$ of
\Bal
  D^\mu F_{\mu\nu}[a] + \zeta \partial_\nu (\partial^\mu a_\mu)
    + \phi \times D_\nu\phi =0,\quad
  D^\mu D_\mu\phi + \LInt_*(\phi)  =0,
\Eal
on an open time interval $(t_0,t_1)$  such that
$\supp(a,\phi)(t)$ is compact for some (and hence all) all $t\in(t_0,t_1)$,
we have
\Beq
 \sup_{t\in(t_0,t_1)} \max
 \br\{
 \|a^i_\mu(t)\|_{H_{k+1}(\rdm)},\
 \|\phi_j(t)\|_{H_{k+1}(\rdm)},\
 \|\partial_t a^i_\mu(t)\|_{H_{k}(\rdm)},\
 \|\partial_t \phi_j(t)\|_{H_{k}(\rdm)}
 \} < \infty
\Eeq
where $i=1,\dots,N^{\fg}$, $\mu=0,\dots,d$, $j=1, \dots,N^\Phi$. Then the
following assertions are valid.

\begin{itemize}
\item[(1)] There exists a unique solution family
\Beqn YMFPUniv
\begin{split}
 \Xi\sol&=(A\sol,\Phi\sol|\ \Lambda\sol,\Ups\sol,\Psi\sol)
\\
 &=\Xi\sol[A\Cau,\dot A\Cau,\Phi\Cau,\dot\Phi\Cau
  |\ \Lambda\Cau,\dot\Lambda\Cau,\Ups\Cau,\dot\Ups\Cau,\Psi\Cau]
 \in \M^{\cE\V}(M\Cau\scinf)
\end{split}
\Eeq
of \eqref{YMGBEq}--\eqref{YMGBEqLast} with the initial data
\Bal
  &(A\sol,\partial_t A\sol,\Phi\sol,\partial_t\Phi\sol
 |\ \Lambda\sol,\partial_t\Lambda\sol,\Ups\sol,\partial_t\Ups\sol,\Psi\sol)(0)
\\
 =\text{ }
 &(A\Cau,\dot A\Cau,\Phi\Cau,\dot\Phi\Cau
 |\ \Lambda\Cau,\dot\Lambda\Cau,\Ups\Cau,\dot\Ups\Cau,\Psi\Cau).
\Eal

\item[(2)] This solution family is universal for the quality
$\cE$, i.~e. any other smooth solution family
$(A',\Phi'|\Lambda',\Ups',\Psi')\in \M^{\cE\V}(Z)$ of
\eqref{YMGBEq}--\eqref{YMGBEqLast}
is in a unique way a pullback of \eqref{YMFPUniv}.

\item[(3)] The image of $M\scinf\sol\seq M\scinf$ is a sub-smf.

\item[(4)] Assume that $\phi\in V^\Phi$ satisfies $\LInt_i(\phi)=0$ for
all $i$ (and hence that $(A,\Phi):=(0,\phi)$ is a solution of the underlying
bosonic equations). Then the superfunctional
\Beq
\begin{split}
 &\Xi\exc_\phi[A\Cau,\dot A\Cau,\Phi\Cau,\dot\Phi\Cau
 |\ \Lambda\Cau,\dot\Lambda\Cau,\Ups\Cau,\dot\Ups\Cau,\Psi\Cau]
\\
 &:=\Xi\sol[A\Cau,\dot A\Cau,\Phi\Cau+\phi\Cau,\dot\Phi\Cau
|\ \Lambda\Cau,\dot\Lambda\Cau,\Ups\Cau,\dot\Ups\Cau,\Psi\Cau]
\end{split}
\Eeq
which lies a priori in $\M^{\cE\V}(M\Cau\scinf)$, restricts to a
superfunctional
\Beq
 \Xi\exc_\phi\in\M^{\cEc\V}(M\Cau).
\Eeq
Moreover, the image of the arising smf morphism $\Xi\exc_\phi: M\Cau\to M$
is again a split sub-smf $M\exc_\phi\seq M$  called the  {\em smf of
excitations around $\phi$}. It has the following universal property:
Given a $Z$-family $\Xi':=(A',\Phi'|\Lambda',\Ups',\Psi')\in\M^{\cEc\V}(Z)$,
the corresponding morphism $\Xi':Z\to M$ factors through $M\exc_\phi$ iff
the $Z$-family $(A',\Phi'+\phi|\Lambda',\Ups',\Psi')\in\M^{\cE\V}(Z)$ is a
solution family.
\end{itemize}
\end{itemize}
\end{thm}

\Brm
(1) The simplest case in which the Theorem applies is the abelian case,
i.~e. that of quantum electrodynamics (where, however, the ghost field is
decoupled from the rest of the system). Here $N^\Phi=0$, and hence
the underlying bosonic theory is free, so that the hypothesis of (ii)
is obviously satisfied.

(2) Unfortunately, this Theorem hangs for the non-abelian case somewhat in
the air, because, at the time being, there is no result which assures
the validity of the hypothesis of (ii). Indeed, the gauge breaking term
extends the dynamics of the underlying bosonic model,
so that the results of Ginebre/Velo and Eardley/Moncrief can no longer
be applied.

Clearly, a bosonic configuration $(a,\phi)\in\Space(M)$ which satisfies
the Lorentz gauge condition
\Beqn Lorentz
 \partial^\mu a_\mu =0
\Eeq
is a solution of the underlying bosonic equations of \eqref{YMGB}
iff it is a solution of the unmodified Yang-Mills equation
\Beqn OrigBYMEq
D^\mu F_{\mu\nu}[a] + \phi \times D_\mu\phi =0;
\Eeq
and, conversely, given a solution of the latter,
one can achieve by a gauge transformation that the constraint \eqref{Lorentz}
is fulfilled. However, a solution of the field equations of
$\cL^{\opn YM-GB }[\ul A]$ which does not satisfy \eqref{Lorentz}, does not
solve \eqref{OrigBYMEq}. Note also that the energy
belonging to \eqref{YMGB} is not positively defined,
since $(\partial_\mu A_0)^2$ enters the Lagrangian, and hence the energy,
with the "wrong" sign.
Thus, there is no guaranty even for boundedness of the $H_1$ norms.

Thus, in the physically most interesting case, the completeness question is
still open. Once it is settled positively, our results will immediately
apply to the standard model, which unifies the Salam-Weinberg electroweak
model with quantum chromodynamics.

(3) Of course, the hypothesis of (4) on $\phi$ is always satisfied for
$\phi:=0$. However, in models with spontanuous gauge breaking, it is from
the beginning more senseful to consider classical solutions around a
minumum of the potential $\LInt$.

(4) In the diagonal gauge, $\zeta=1$, we have modulo a total derivative
\Beq
 L\kin[\ul A] \equiv -\frac12 \partial_\mu\ul A_\nu \partial^\mu\ul A^\nu,
\Eeq
i. e. it describes $dm$ scalar fields, and we may use a straightforward
reduction to first order, with ascribing to the $A_\mu$ the smoothness
offsets $1$.
Unfortunately, this ceases to be true for arbitrary $\zeta\not=0$, and
we face a situation analogous to that arising in the temporal gauge:
The kinetic operator arising from \eqref{YMGB} is
\Beq
 K^{\mu\nu} = \delta^{\mu\nu} \KlGord + (\zeta-1)\partial^\mu\partial^\nu.
\Eeq
Setting $p:=(0,p_1,\dots,p_d)\trp$, $e:=(1,0,\dots,0)\trp$ and
\Beq
 \alpha = \i(\zeta-1)( \frac1\zeta e p\trp - p e\trp),\quad
 \beta =  p^2\cdot1_{(d+1)\times(d+1)} + p^2 \frac{1-\zeta}\zeta ee\trp
 + (\zeta-1)pp\trp,
\Eeq
the influence function $\cA$ is given by
$\hat\cA(t,p) =(2\pi)^{-d/2}1_{m\times m}\otimes\hat\cA(t,p)$
where $\hat\cA(t,p)$ is determined by
\Beq
 \frac{d^2}{dt^2} \hat\cA(t,p) + \alpha \frac d{dt}\hat\cA(t,p) +
 \beta\hat\cA(t,p) =0,\qquad
 \hat\cA(0,p)=0,\quad \frac d{dt} \hat\cA(0,p)=1_{(d+1)\times(d+1)}.
\Eeq
The solution is given by
\Beq
 \hat\cA(t,p)= g_0(t,p)(1 - ee\trp - \frac1{p^2} pp\trp)
 + g_{ee}(t,p) ee\trp
 + \frac{g_{ep}(t,p)}{\br|p|} (ep\trp - \frac1\zeta pe\trp)
 + \frac{g_{pp}(t,p)}{p^2}pp\trp
\Eeq
with
\Beq
\begin{array}{ll}
 g_0(t,p) = \displaystyle \frac{\sin(\br|p|t)}{\br|p|},\quad
 &g_{ee}(t,p) = \displaystyle \frac{(1 - \zeta)}2  t \cos (\br|p|t) +
  \frac{(1 + \zeta)}{2\br|p|} \sin (\br|p| t)
 ,\\
 &\\
 g_{ep}(t,p) = \displaystyle \frac{\i(\zeta-1)}2  t \sin(\br|p|t),\quad
 &g_{pp}(t,p) = \displaystyle \frac{\zeta-1}{2\zeta} t \cos (\br|p|t) +
 \displaystyle \frac{\zeta+1}{2\zeta\br|p|} \sin (\br|p| t).
\end{array}
\Eeq
Unfortunately, the terms proportional to $\zeta-1$ spoil the decay of
$\hat\cA(t,p)$ with $p\to\infty$, i. e. we cannot assign to the $A_\mu$
the smoothness degree $1$. Therefore, we will have to use a more
complicated reduction to first order, as in \eqref{TG1stOrder}.
\Erm

\subsection{The first-order system}
Besides the magnetic field strengthes \eqref{DefBab}, we will use
the electric field strengthes
\Beqn DefEa
 E_a=\partial_0 A_a - \partial_a A_0 +\br[A_0,A_a].
\Eeq
Introducing also the longitudinal component
\Beqn DefL
 L=\partial^\mu A_\mu,
\Eeq
\eqref{YMGBEq} takes the form
\Baln DaEa
 &-D^aE_a + \zeta\partial_0L + J_0 =0,
\\ \label{D0Ea}
 &D^0E_b + D^aB_{ab} + \zeta\partial_bL + J_b=0.
\Eal
The computation of the covariant divergence of the matter current
is similar as in the derivation of \eqref{DtCst}, using this time
the matter field equations \eqref{EvYMAgh}, \eqref{YMFPPsi},
\eqref{YMGBEqLast}; we find for any solution family
\Beqn DmuJmu
 D^\mu J_\mu
 = - \i\br[D^\mu\partial_\mu\Lambda, \Ups] - \i\br[\partial_\mu\Lambda,
D^\mu\Ups]
 = - \i \br[\partial_\mu\Lambda, D^\mu\Ups].
\Eeq

Taking the covariant derivative $D^\nu$ of \eqref{YMGBEq} we get, using
\eqref{DmuJmu}, an evolution equation for $L$:
\Beqn EvLong
 \zeta D^\mu\partial_\mu L=\i\br[\partial_\mu\Lambda, D^\mu\Ups].
\Eeq
As in \eqref{TG1stOrder}, we forget the definition \eqref{DefBab} of the
magnetic field strengthes and consider them as independent antisymmetric
tensor field $B_{ab}$. Also, we will use for convenience the $E_a$ instead
of $\dot A_a$. Finally, we introduce again
a new $\fg$-valued scalar field $L$ for the  longitudinal component.
Thus, we consider the following system in the commuting fields
$A_0,\dot A_0,A_a,E_a,B_{ab},L,\dot L,\Phi,\dot\Phi$ and the
anticommuting fields $\Lambda,\dot\Lambda,\Ups,\dot\Ups,\Re\Psi,\Im\Psi$:
\pagebreak[2]
\begin{subequations}
\Baln ExtGBSyst
 \partial_t A_0 &= \dot A_0, &&
 \\\label{DtdA0}
 \partial_t \dot A_0 &
 = \Delta A_0 + (\zeta-1)\dot L - \br[A_0,L+\dot A_0]-\br[\partial_aA_0,  A_a]
 \\
 &\quad - \br[A_a,  E_a] + J_0
 &&\text{(see below),}\notag
 \\\label{DtAa}
 \partial_t A_a &= E_a + \partial_aA_0 - \br[A_0, A_a]
 &&\text{(by \eqref{DefEa}),}
 \\ \label{DtEa}
 \partial_t E_a &=\partial_bB_{ba} - \br[A_0,E_a]
    + \br[A_b,B_{ba}] + \zeta\partial_a L + J_a
 &&\text{(by \eqref{D0Ea}, \eqref{FPMCur}),}
 \\
 \label{TDConstr}
 \partial_t B_{ab} &=\partial_a E_b - \partial_b E_a - \br[A_0,B_{ab}]
 + \br[A_a,E_b] -\br[A_b,E_a]
  && \text{(Bianchi id.),}
 \\
 \partial_t L&=\dot L, &&
 \\\label{DtdotL}
 \partial_t\dot L&= \Delta L + \br[A_a,\partial_aL] - \br[A_0,\dot L]
 \\
 &\quad + \frac\i\zeta \left(\br[\dot\Lambda, \dot\Ups + {\br[A_0,\Ups]}]
  - \br[\partial_a\Lambda, \partial_a\Ups + {\br[A_a,\Ups]}] \right)
 &&\text{(by \eqref{EvLong}),} \notag
 \\
 \partial_t\Phi &=\dot\Phi, &&
 \\\label{DtdotPhi}
 \partial_t\dot\Phi&=\Delta\Phi + L\times\Phi
 - 2A_0\times\dot\Phi + 2A_a\times\partial_a\Phi
 \\
 &\quad - A_0\times(A_0\times\Phi) + A_a\times (A_a\times \Phi) + \LInt_*(\Phi)
 && \text{(by \eqref{YMGBEqLast}),}\notag
 \\
 \partial_t\Lambda&= \dot\Lambda,
 \\\label{DtdotAgh}
 \partial_t\dot\Lambda&= \Delta\Lambda + \br[A_a,\partial_a\Lambda]
   - \br[A_0,\dot\Lambda]
 && \text{(by \eqref{EvYMAgh}),}
 \\
 \partial_t\Ups&= \dot\Ups,
 \\\label{DtdotUps}
 \partial_t\dot\Ups&= \Delta\Ups + \br[L,\Ups] - \br[A_0,\dot\Ups]
   + \br[A_a,\partial_a\Ups]
 && \text{(by \eqref{EvYMUps}),}
 \\\label{ExtGBSystLast}
 \partial_t\Psi &= -(\gamma^0)^{-1} \br(\gamma^a \partial_a \Psi
   + \gamma^a A_a\times\Psi  +  m^\Psi\Psi) - A_0 \times\Psi
 &&\text{(by \eqref{EvYMPsi}).}
\Eal
\end{subequations}
where we have abbreviated
\Bal
  J_0&:=-\i\br[\dot\Lambda,\Ups]
  + \Phi\times(\dot\Phi + A_0\times\Phi) - \i\dcj\Psi\times\gamma_0\Psi,
 \\
 J_a&:=-\i\br[\partial_a\Lambda,\Ups]
  + \Phi\times (\partial_a\Phi + A_a\times \Phi)
  - \i\dcj\Psi\times\gamma_a\Psi.
\Eal
For the derivation of \eqref{DtdA0} we note that \eqref{DaEa} yields
\Bal
 -\partial^a(\partial_0 A_a - \partial_a A_0 +\br[A_0,A_a]) - \br[A^a,  E_a]
 + \zeta\partial_0L + J_0 &= 0,
\\
 -\partial_0 \partial^aA_a + \Delta A_0 - \br[A_0, \partial^a A_a]
 - \br[\partial^aA_0, A_a]  - \br[A^a,  E_a] + \zeta\partial_0L + J_0 &= 0,
\\
 -D_0 (L + \partial_0A_0) + \Delta A_0  -\br[\partial^aA_0, A_a]
 - \br[A^a,  E_a] + \zeta\partial_0L + J_0 &= 0,
\Eal
which now rewrites to \eqref{DtdA0}.

{}From \eqref{DefL}, \eqref{DaEa} we get constraints for $L,\dot L$:
\Baln ConstL
 L&= - \dot A_0 + \partial_a A_a,
 \\
 \label{ConstDL}
 \dot L &= \frac{1}{\zeta}\left(\partial_aE_a + \br[A_a,E_a] - J_0\right)
\Eal
(note that trying to use the latter relation as evolution equation would
result in smoothness problems).

Now the system \eqref{ExtGBSyst}--\eqref{ExtGBSystLast} together with
the constraints  \eqref{DefBab}, \eqref{ConstL},  \eqref{ConstDL}
is equivalent with the original system \eqref{YMGBEq}--\eqref{YMGBEqLast}:

\begin{lem}\label{YMFPLem}
(i)
A configuration family $(A,\Phi|\Lambda,\Ups,\Psi)$ is a solution family of
\eqref{YMGBEq}--\eqref{YMGBEqLast} iff
\Beqn ExtConfFam
 (A_0,\partial_tA_0,A_a,E_a,B_{ab},L,\partial_tL,\Phi,\partial_t\Phi|
 \Lambda,\partial_t\Lambda,\Ups,\partial_t\Ups,\Re\Psi,\Im\Psi)
\Eeq
where  $E_a$, $B_{ab}$, and $L$ are defined by
\eqref{DefEa}, \eqref{DefBab},  and \eqref{DefL}, respectively,
is a solution family of the system \eqref{ExtGBSyst}--\eqref{ExtGBSystLast}.
In that case, the constraints \eqref{ConstL}, \eqref{ConstDL} are satisfied.

(ii) Conversely, let given a solution family \eqref{ExtConfFam} of
\eqref{ExtGBSyst}--\eqref{ExtGBSystLast} which satisfies the constraints
\eqref{DefBab}, \eqref{ConstL},  \eqref{ConstDL} at $t=0$. Then these are
satisfied at all times, \eqref{DefL} holds, and $(A,\Phi|\Ups,\Psi)$ will
be a solution family of \eqref{YMGBEq}--\eqref{YMGBEqLast}.

(iii) The system  \eqref{ExtGBSyst}--\eqref{ExtGBSystLast} belongs to the
class considered in \cite{[CAUCHY]}, \cite{[ABSEVEQ]}, with assigning the
smoothness offset $1$ to $A_0,\Ups,\Phi,L$, and $0$ to all other components.
Moreover, the system  \eqref{ExtGBSyst}--\eqref{ExtGBSystLast} is causal in
the sense of \cite{[CAUCHY]}, \cite{[ABSEVEQ]}.
\end{lem}

\Brm
The Lemma and the results of \cite{[ABSEVEQ]} yield a short-time solvability
result for \eqref{YMGBEq}--\eqref{YMGBEqLast}.
\Erm

\begin{proof}
Ad (i). This is clear by construction.
Ad (ii).
\eqref{YMFPPsi} is clear. \eqref{DtAa} shows that \eqref{DefEa} is satisfied.
Now the Bianchi identity \eqref{Bianchi} yields
\Beq
\partial_t (B_{ab}-F_{ab}[A]) =\br[A_0,B_{ab}-F_{ab}{[A]}].
\Eeq
Therefore, the vanishing of the initial value implies $B_{ab}=F_{ab}[A]$,
i.~e. the validity of \eqref{DefBab}.

Now we have to reconstruct $L$. Setting for shortness
$\lambda:=\partial_aA_a - \dot A_0 - L$, we
get from \eqref{DtdA0} and \eqref{DtAa}
\Bal
 \partial_t\lambda
 &=\partial_a(E_a + \partial_aA_0 - \br[A_0, A_a] ) -\Delta A_0 - \zeta\dot L
 + \br[A_0,L + \dot A_0] + \br[\partial_aA_0,  A_a] + \br[A_a,  E_a]  - J_0
\\
 &= \partial_aE_a  + \br[A_0, \partial_aA_a]
 - \zeta\dot L + \br[A_0,L + \dot A_0] + \br[A_a,  E_a] - J_0
\Eal
i.~e.
\Beqn D0Lambda
 D_0\lambda=D_aE_a  - \zeta\dot L   - J_0
\Eeq
(essentially, this is the derivation of \eqref{DtdA0} read backwards).
Hence, using \eqref{DtEa},
\Bal
 D_0D_0\lambda
 &= D_a D_0E_a  - D_0 (\zeta \dot L + J_0)
 \\
 &= D_a D_bB_{ba} + D_a(\zeta\partial_a L + J_a) - D_0 (\zeta \dot L + J_0)
  = \zeta D^\mu\partial_\mu L + D^\mu J_\mu.
 \notag
\Eal
For evaluating the r.h.s., we rewrite the matter field equations given by
\eqref{DtdotL}, \eqref{DtdotPhi}, \eqref{DtdotAgh}, \eqref{DtdotUps},
\eqref{ExtGBSystLast} to
\Bal
 &\zeta D^\mu\partial_\mu L - \i\br[\partial^\mu \Lambda,D_\mu\Ups]=0,
 &&\partial^\mu D_\mu\Ups - \br[\lambda,\Ups]=0,\qquad
 D_\mu\partial^\mu \Lambda =0,
 \\
  &\br(\gamma^\mu D_\mu + m^\Psi)\Psi =0,
  &&D^\mu D_\mu\Phi + \LInt_*(\Phi)  - \lambda\times\Phi=0.
\Eal
With a similar computation as in the derivation of \eqref{DtCst}, we get
\Beqn D0D0Lambda
 D_0D_0\lambda = \Phi\times(\lambda\times\Phi).
\Eeq
Now the constraint \eqref{ConstL} at $t=0$ says that $\lambda(0)=0$, and
\eqref{ConstDL} at $t=0$ says that the r.~h.~s. of \eqref{D0Lambda}
vanishes at $t=0$, so that $\partial_t\lambda(0)=0$. Therefore
\eqref{D0D0Lambda} now implies $\lambda=0$ for all times, i.~e.
\eqref{ConstL} is satisfied for all times.
Now \eqref{D0Lambda} implies the validity of \eqref{ConstDL} for all times
as well as \eqref{DaEa}, while \eqref{DtEa} yields \eqref{D0Ea}. The
remaining equations are now obviously satisfied.

Ad (iii).
For finding the influence function $\hA$ of the free evolution, we may
consider again the sectors separately: We have
\Beq
 \hA=\opn diag (\hA^{AEBL},\hA^\Phi,\hA^\Lambda,\hA^\Ups,\hA^\Psi).
\Eeq
Here $\hA^\Lambda,\hA^\Ups$ are both given by \eqref{IAPhi} with $m^\Phi=0$,
\ \ $\hA^\Phi$ is given by \eqref{IAPhi} with
$m^\Phi_{ij}:= \frac{\partial^2}{\partial\ul\Phi_i\partial\ul\Phi_j}\LInt(0)/2$
again,  and $\hA^\Psi=A^\Psi$ is again the same as in \ref{Yukawa}.

The linearized evolution of both $L$ and $A_0$ is given by the massless
Klein-Gordon equation, with $A_0$ having $\dot L$ as source term. Also,
the linearized evolution of $A_a,E_a,B_{ab}$ is the same as in
the proof of Lemma \ref{LemTmpGaugeEv},
apart from the source terms containing $A_0$ and $L$;
the latter affect only the last two columns of the matrix below.
These observations lead to the following structure for
the influence function for $A_0,\dot A_0,A_a,E_a,B_{ab},L,\dot L$:
\Beqn hAEBL
\hA^{AEBL} =
\begin{pmatrix}
 \partial_t\hP_0    &  \hP_0           &0 & 0 & 0     &\hX^L_0
  & \hX^{\dot L}_0 \\
 \partial_t^2\hP_0&\partial_t\hP_0&0 & 0 & 0     &\partial_t{\hX}^L_0
 &\partial_t{\hX}^{\dot L}_0 \\
 \i p_c\hP_0  &\i p_c\int^t_0d\tau\hP_0(\tau,p) &   &   &  &\hX^L_c
 & \hX^{\dot L}_c \\
 0                       &   0              &&\hA^{AEB}  &&\hA^L_c
 & \hA^{\dot L}_c \\
 0                       &   0              &   &   &  &\hA^L_{cd}
 & \hA^{\dot L}_{cd} \\
 0                       &   0              &0 & 0 & 0 &\partial_t\hP_0
 & \hP_0 \\
 0                       &   0              &0 & 0 & 0 &\partial_t^2\hP_0
 & \partial_t\hP_0 \\
\end{pmatrix}
\Eeq
where $\hA^{AEB}$ is as in \eqref{hAAEB}, \ \
$\hP_0:=(2\pi)^{-d/2}\sin\br|p|t/\br|p|$, and
the first five lines in the last two columns are yet to be determined.
Letting $I$ stand for either $L$ or $\dot L$, they are determined by
\Bal
 &\hX^I_0(0)=\partial_t\hX^I_0(0)= \hX^I_c(0)=\hA^I_c(0)=\hA^I_{cd}(0)=0,
\\
 &\partial_t^2\hX^I_0 = -p^2\hX^I_0 + (\zeta-1) \times
 \begin{cases}
   \partial_t^2\hP_0& \text{for $I=L$} \\
   \partial_t\hP_0  &  \text{for $I=\dot L$}
  \end{cases},
\\
 &\partial_t\hX^I_c = \hA^I_c + \i p_c \hX^I_0,
 \quad\quad
 \partial_t\hA^I_{cd} = \i p_c\hA^I_d - \i p_d\hA^I_c ,
\\
 &\partial_t\hA^I_c = \i p_d\hA^I_{dc} + \i\zeta p_c
 \begin{cases}
   \partial_t\hP_0 & \text{for $I=L$} \\
   \hP_0           & \text{for $I=\dot L$}
  \end{cases}.
\Eal

Abbreviating $C:=\cos\br|p|t$, \ \ $S:=\sin\br|p|t$, and omitting
the colour indices, one checks that the solution is given by
$(2\pi)^{d/2}\hA^{AEBL}=$
\Beq
\begin{pmatrix}
 C  &\frac{1}{\br|p|}S  &\dots
    &\frac{\zeta-1}{2\br|p|} \br(\br|p|tC - S)
    &\frac{\zeta-1}{2\br|p|} tS
 \\
 -\br|p|S &C     &\dots
   &\frac{(1-\zeta)}2\br|p|tS
   &\frac{\zeta-1}{2\br|p|}\br(S + t\br|p|C)
 \\
 \frac{\i}{\br|p|}p_cS
 &\frac{\i \br(1-C) }{p^2} p_c
 &\dots
   &\frac{\i p_c}{2\br|p|^2} \br({-2C +(\zeta-1)\br|p|tS})
   &\frac{\i p_c}{2\br|p|^3}
    \br({ \br|p|t(2\zeta + (1-\zeta)C) - (1+\zeta)S })
 \\
 0                   &0 &\dots
   &\frac{\i\zeta}{\br|p|}p_cS
   &\frac{\i\zeta}{\br|p|^2}p_c\br(1-C)
 \\
 0                   &0 &\dots &0        &0
 \\
 0                   &0 &\dots &C        & \frac{1}{\br|p|}S
 \\
 0                   &0 &\dots &-\br|p|S &C
\end{pmatrix}
\Eeq
where the three middle columns, which are already known from
\eqref{hAEBL}, \eqref{hAAEB}, have been omitted.

The assertion (iii) is now easy to check.
\end{proof}

\begin{proof}[Proof of Thm. \ref{YMFPThm}]
Ad (i).
This follows from Lemma \ref{YMFPLem} and the results of \cite{[ABSEVEQ]}.

Ad (ii). Let
\Beq
 V\ext := \fg\otimes\br(\mathbb R \oplus \mathbb R \oplus \rdm \oplus \rdm
 \oplus \Lambda^2\rdm \oplus \mathbb R \oplus \mathbb R)
 \oplus V^\Phi \oplus V^\Phi
 \ \oplus \ \Pi(\fg\oplus \fg\oplus \fg\oplus \fg)
 \oplus \Pi(V^\Psi\otimes_{\mathbb C} V\D)
\Eeq
be the field target space for the extended system
\eqref{ExtGBSyst}--\eqref{ExtGBSystLast}. Thus, the corresponding smf of
smooth Cauchy data is
$M^{\opn extCau }\scinf=\L(\cE\Cau\otimes V\ext)$. Consider the smf morphism
\Beqn MCau2MExtCau
 M\Cau\scinf\too{\quad(A_0\Cau,\dot A_0\Cau,A_a\Cau,E_a\Cau,B_{ab}\Cau,
 L\Cau,\dot L\Cau,\Phi\Cau,\dot\Phi\Cau|
 \Lambda\Cau,\dot\Lambda\Cau,\Ups\Cau,\dot\Ups\Cau,\Psi\Cau)\quad}
 M^{\opn extCau }\scinf,
\Eeq
where
\Bal
 E_a\Cau&:=\dot A_a\Cau - \partial_a A_0\Cau +\br[A_0\Cau,A_a\Cau],\quad
 B_{ab}\Cau:= \partial_a A\Cau_b - \partial_b A\Cau_a +\br[A\Cau_a,A\Cau_b],
 \\
 L\Cau&:= - \dot A_0\Cau+\partial^a A_a\Cau,
 \\
 \dot L\Cau &:=  \frac{1}{\zeta}\Bigl(\i\br[\dot\Lambda\Cau,\Ups\Cau]
 - \Phi\Cau\times(\dot\Phi\Cau + A_0\Cau\times\Phi\Cau)
 + \i\dcj\Psi\Cau\times\gamma_0\Psi\Cau
 \\
 &\quad + \partial_aE_a\Cau + \br[A\Cau_a,E\Cau_a]\Bigr).
\Eal
Letting $Y$ denote temporarily the image of its underlying map,
the hypothesis just says that the system
\eqref{ExtGBSyst}--\eqref{ExtGBSystLast}
is $(\cH^{\opn V \ext_k},Y\cap \cH^{\opn V \ext_k})$-complete, where we use the
Sobolev space $\cH^{\opn V \ext_k}$ as defined in \cite{[CAUCHY]},
\cite{[ABSEVEQ]}. Hence, by \cite[Thm. 3.4.3]{[CAUCHY]}, there exists a
solution family
\Bml
 \Xi\extsol
 =(A_0\extsol,\dot A_0\extsol, A_a\extsol,E_a\extsol,B_{ab}\extsol,
   L\extsol,\dot L\extsol,\Phi\extsol,\dot\Phi\extsol|
  \\
  \Lambda\extsol,\dot\Lambda\extsol,\Ups\extsol,\dot\Ups\extsol,
  \Psi\extsol)\in \M^{\cE\otimes V\ext}(M\Cau\scinf)
\Eml
which has \eqref{MCau2MExtCau} as its Cauchy data. Now
$\Xi\sol:=(A\extsol,\Phi\extsol|\Lambda\extsol,\Ups\extsol,\Psi\extsol)$
will be the universal solution family wanted. For showing that its image
is a sub-smf, the proof of \cite[Thm. 2.4.1]{[ABSEVEQ]} carries over again.
\end{proof}


\hfill

{\em Author's address:} Ackerstra\ss e 11, 10 115 Berlin, Germany
\quad(ts@berlin.snafu.de)

\end{document}